\documentclass[final,3p,times]{elsarticle}

\usepackage{tabularx}
\usepackage{hyperref}
\usepackage{amsmath}
\usepackage{amsfonts}
\usepackage{amsthm}
\usepackage{graphicx}
\usepackage{subcaption}
\usepackage{amssymb}
\usepackage{color}

\DeclareSymbolFont{CMMI}{OML}{ccm}{m}{it}
\DeclareMathSymbol{v}{\mathalpha}{CMMI}{"76}

\newtheorem{proposition}{Proposition}

\theoremstyle{definition}

\renewcommand{\v}[1]{\ensuremath{\mbox{\boldmath$ #1 $}}} 
 
\newcommand{\uv}[1]{\ensuremath{\mathbf{\hat{#1}}}} 
\newcommand{\pd}[2]{\frac{\partial #1}{\partial #2}} 
 
\newcommand{\pddo}[3]{\frac{\partial^2 #1}{\partial #2 \partial #3}} 
\renewcommand{\div}[1]{\nabla \cdot #1} 
\newcommand{\coll}[1]{\left(\pd{ #1}{t}\right)_c} 
\newcommand{\dxdv}{\thinspace dx\thinspace dv}

\newcommand{\vte}{v_{th,e}}

\DeclareSymbolFont{CMBI}{OML}{ccm}{b}{it}
\DeclareMathSymbol{\vv}{\mathalpha}{CMBI}{"76}
\newcommand{\vi}{v_i}
\newcommand{\vj}{v_j}

\newcommand{\vu}{\v{u}}

\newcommand{\divv}[1]{\nabla_{\vv} \cdot #1}
\newcommand{\vE}{\v{E}}
\newcommand{\vB}{\v{B}}

\newcommand{\vcen}{\check{v}}

\newcommand{\wpump}{\omega_{\text{pump}}}

\newcommand{\omegace}{\Omega_{ce}}

\newcommand{\ignore}[1]{}  

\DeclareMathAlphabet{\mathpzc}{OT1}{pzc}{m}{it}

\newcommand{\incfig}{\centering\includegraphics}
\setkeys{Gin}{width=0.9\linewidth,keepaspectratio}

\newcommand{\eqr}[1]{Eq.\thinspace(#1)}
\newcommand{\eqsr}[2]{Eqs.\thinspace(#1)-(#2)}

\newcommand{\pfrac}[2]{\frac{\partial #1}{\partial #2}}
\newcommand{\pfraca}[1]{\frac{\partial}{\partial #1}}

\newcommand{\gvec}[1]{\boldsymbol{#1}}

\newcommand{\gvs}{\nabla_{\vv}}

\usepackage{graphicx}
\usepackage{color}

\usepackage{tabularx}

\newcommand{\comment}[1]{\textit{\textcolor{red}{#1}}}
\renewcommand{\comment}[1]{}

\setkeys{Gin}{width=0.9\linewidth,keepaspectratio}

\newcommand{\gke}{{\tt Gkeyll}}

\newcommand{\cramplist}{
	\setlength{\itemsep}{0in}
	\setlength{\partopsep}{0in}
	\setlength{\topsep}{0in}}

\journal{Journal of Computational Physics}

\begin{document}

\begin{frontmatter}

\title{Conservative Discontinuous Galerkin Schemes for Nonlinear Fokker-Planck Collision Operators}
\author[pppl]{A. Hakim}
\author[mit,dart]{M. Francisquez}
\author[umd]{J. Juno}
\author[pppl]{G.~W. Hammett}
\address[pppl]{Princeton Plasma Physics Laboratory, Princeton, NJ 08543-0451}
\address[mit]{MIT Plasma Science and Fusion Center, Cambridge, MA, 02139}
\address[dart]{Department of Physics and Astronomy, Dartmouth College, Hanover, NH, 03755}
\address[umd]{IREAP, University of Maryland, College Park, MD, 20742}

\begin{abstract}
We present a novel discontinuous Galerkin algorithm for the solution of a class of Fokker-Planck collision operators. These operators arise in many fields of physics, and our particular application is for kinetic plasma simulations. In particular, we focus on an operator often known as the ``Lenard-Bernstein,'' or ``Dougherty,'' operator. Several novel algorithmic innovations are reported. The concept of weak-equality is introduced and used to define weak-operators to compute primitive moments needed in the updates. Weak-equality is also used to determine a reconstruction procedure that allows an efficient and accurate discretization of the diffusion term. We show that when two integration by parts are used to construct the discrete weak-form, and finite velocity-space extents are accounted for, a scheme that conserves density, momentum and energy exactly is obtained. One novel feature is that the requirements of momentum and energy conservation lead to unique formulas to compute primitive moments. Careful definition of discretized moments also ensure that energy is conserved in the piecewise linear case, even though the $v^2$ term is not included in the basis-set used in the discretization. A series of benchmark problems are presented and show that the scheme conserves momentum and energy to machine precision. Empirical evidence also indicates that entropy is a non-decreasing function. The collision terms are combined with the Vlasov equation to study collisional Landau damping and plasma heating via magnetic pumping. We conclude with an outline of future work, in particular with some indications of how the algorithms presented here can be extended to use the Rosenbluth potentials to compute the drag and diffusion coefficients.
\end{abstract}

\end{frontmatter}

\section{Introduction} \label{sec:introduction}

The motion of charged particles in a plasma is due to self-consistent mean electromagnetic field and cumulative effects of fluctuating fields. The latter can be approximated effectively as small-angle collisions. This approximation is in contrast to the case of neutral particles that travel in straight lines between ``hard-sphere'' collisions. A fully ionized plasma is described by the Vlasov-Maxwell-Fokker-Planck (VM-FP) equation for the phase-space distribution function, $f(t,\v{x},\vv)$, written as
\begin{align} \label{eq:boltzmann}
\pd{f_s}{t}+\div{\left(\vv f_s\right)}+\divv{\left(\v{a}_s f_s\right)} = \coll{f_s}.
\end{align}
Here $\v{a}_s=\left(q_s/m_s\right)\left(\vE+\vv\times\vB\right)$ is the acceleration due to the Lorentz force, and $q_s$ and $m_s$ the charge and mass of species $s$ respectively. The mean electric and magnetic fields, $\v{E}$ and $\v{B}$ are determined from net charges and currents generated due to the motion of the particles, and are governed by the Maxwell's equations of electromagnetism. The effect of small-angle collisions, written on the right hand side of the above equation, is described by the nonlinear Fokker-Planck operator (FPO)~\cite{Rosenbluth1957} as
\begin{align}\label{eq:FProsenbluth}
\coll{f_s} = C[f] = -\pd{}{\vi}\left(\left\langle \Delta \vi\right\rangle_s f_s\right)+\frac{1}{2}\pddo{}{\vi}{\vj}\left(\left\langle\Delta \vi\Delta \vj\right\rangle_s f_s\right).
\end{align}
Here $\langle \Delta\vi \rangle_s$ and $\langle \Delta \vi\Delta \vj \rangle_s$ are average increments per unit time due to the effect of collisions governed by the inverse-square force between individual particles. In general, the increments are determined either from a Landau integral formulation or from the Rosenbluth potentials that themselves are computed by solving elliptic boundary value problems involving the distribution function. This results in a nonlinear integro-differential equation for the effect of collisions on the evolution of the distribution function. See equations (17)--(20) in~\cite{Rosenbluth1957} or Refs.~\cite{2005ctmp.book,Dwight-Nicholson83,Landau1936} for details. We refer to this specific system as the Rosenbluth (Rosenbluth/Landau) Fokker-Planck operator to avoid confusion. We also note that the Landau form of the Fokker-Planck operator is mathematically equivalent to the Rosenbluth form, and is often prefered to prove various conservation properties and the H-theorem of the operator.

In this paper we instead approximate the increments from self-collisions as $\langle \Delta v_i \rangle_s = -\nu_{ss} (v_i-u_{i,s})$ and $\langle \Delta v_i \Delta v_j \rangle_s = 2\nu_{ss} v_{th,s}^2\delta_{ij}$, where $\nu_{ss}$ is the collision frequency, the primitive moments $\v{u}_s$ and $v_{th,s} = \sqrt{T_s/m_s}$ are the mean velocity and thermal speed of the particles, respectively. The full approximation, including cross-species collisions is given in the next section.  Clearly, this is a significant simplification. However, the essential features of the complete system are still present: the equation is a nonlinear integro-differential equation as the mean velocity $\v{u}_s$ and the thermal speed $v_{th,s}$ are determined by moments of the distribution function; the effect of both drag and diffusion on the particle distribution function is included. The diffusion term means that fine scale oscillations in velocity space are damped more quickly than long-scale oscillations, as in the full Rosenbluth/Landau Fokker-Planck operator, and unlike the algebraic Krook operator that damps all scales at the same rate. One of the weakness of this simplification is in the use of a mean collision frequency independent of the particle velocity. In the full system the effective collision frequency is smaller for higher speed particles and goes like $1/v^3$. It is possible to improve this model by making the collision frequency a function of velocity, but the definitions of the coefficients $u_{i,s}$ and $v_{th,s}^2$ must be modified to preserve momentum and energy conservation. This deficiency means that in our simplified formulation high energy tails in the distribution function are equally impacted by collisions as the low energy bulk, and this can result in inaccurate physics in some situations. 

The simplified FPO already illustrates the challenges in designing a robust, efficient, and production quality numerical method that works well with the discretization of the collisionless terms. Conservation in the presence of both drag and diffusion and finite velocity extents is challenging to maintain. Moments need to be self-consistently determined with the discretization of the collision operator. Diffusion terms need to be handled carefully to ensure momentum and energy conservation. In fact, many features of the algorithm designed to handle these challenges remain essentially unchanged for the full operator, except that now the increments need to be determined from elliptic equations in velocity space. Solving for the Rosenbluth potentials adds additional difficulties that will be addressed in a future work that builds on the scheme presented here.

Fokker-Planck operators arise in many fields of physics, besides the plasma case presented above. Hence, versions of the scheme presented here can be more broadly applied. For example, Fokker-Planck equations are used in studying Brownian motion, Ornstein-Uhlenbeck process, self-gravitating systems~\cite{Padmanabhan:1990wk}, modeling neuronal dynamics in the brain~\cite{Buice2013}. Fokker-Planck equations are used to model stock prices as random walks, and are widely used in the Black-Scholes approach to pricing financial instruments. For a comprehensive overview of the equation and its application see the text of Risken~\cite{Risken89} or Gardner\cite{Gardner2009}.

Our approach to discretize the Vlasov-Maxwell-Fokker-Planck operator is to use a version of the discontinuous Galerkin (DG) scheme. Previously~\cite{Juno2018}, we presented a nodal DG scheme to solve the collisionless system. This scheme is robust and efficient, and was recently modified to use modal basis functions and computer algebra system (CAS) generated code to further enhance its efficiency. This modification will be reported in a separate publication~\cite{Hakim2018}. Here we use the same technique, that is, a modal DG scheme combined with CAS generated code to design a scheme that conserves number density, momentum density and energy. Along with an energy conserving scheme for the collisionless terms, we obtain a robust energy conserving scheme for the full Vlasov-Maxwell-Fokker-Planck equation.

Although we focus on the DG scheme, previous work on numerical integration of kinetic equations has employed finite difference~\cite{Idomura2008}, finite volume~\cite{Dorf2012} and spectral methods~\cite{Jorge2017}. Each has advantages for specific applications. In particular~\cite{Taitano:2015eb} presents an implicit method to discretize the multi-species Fokker-Planck operator, using a Jacobian-free Newton-Krylov scheme. Their scheme exactly conserves mass, momentum and energy. Although the scheme presented here is explicit, implicit methods, either via acceleration or inversion of linear systems will be explored in the future. An explicit collision operator is easier to integrate into the Vlasov-Maxwell system as a first step, and there are a number of applications of interest where the collision frequency is relatively weak so an explicit method is fine. In~\cite{Hirvijoki:2017ei} the Landau form of the FPO is discretized using a Galerkin formulation and, alternatively in~\cite{Hirvijoki:2017ei,Hirvijoki:2018ti}, a direct discretization of the ``metriplectic'' structure of the system. There are some advantages of discretizing the Landau form, as done in~\cite{Hirvijoki:2017ei,Hirvijoki:2018ti}, as the proofs of conservation and H-theorem are direct and can be exploited in constructing the scheme. However, the Landau form involves convolutions in 6D velocity space, making the scheme expensive. 

In fusion physics there is a long history of methods developed to solve the Rosenbluth/Landau-Fokker-Planck operator. This was driven by the need to understand the impact of collisions on magnetic confinement and heating.  In~\cite{McCoy:1981fv,HammettThesis} and subsequently in the book~\cite{Killeen1986} a method to solve the bounce-averaged collision operator, including a quasi-linear radio-frequency (RF) term is developed. The essential idea there~\cite{McCoy:1981fv,HammettThesis} was to expand the distribution function in spherical harmonics and use the equations (combined with finite-differences) for the coefficients to update the collision term. This early work has resulted in a widely used Fokker-Planck code called CQL3D (see \url{https://www.compxco.com/cql3d.html}). This code also includes relativistic effects and has been used to study, among other things, runaway electron physics. Note that in these approaches the focus is on the detailed effects of collisions and usually the collisionless terms (streaming and Lorentz forces) are not included, or treated only approximately. More recently, in~\cite{Hager:2016gq} a conservative finite-volume solver was developed for the Landau form of the operator, for use in a edge gyrokinetic particle-in-cell code, leading to a hybrid continuum/particle-in-cell scheme for the collisional gyrokinetic equations. (These are just some examples, there are many others.  All of the major gyrokinetic codes used in fusion research include implementations of Fokker-Planck collision models of various forms, see for example \cite{2017PPCF...59d5005B,Barnes2009,Pan2019,NAKATA201561}.)

High-order discontinuous Galerkin (DG) schemes have many advantages. High order solutions obtained from a DG scheme can offer increased accuracy at reduced cost. In addition, the data locality of a DG scheme is an attractive feature for high performance computing on modern processor architectures that prefer intense computations over data communication. The locality and flexibility of DG also allows constructing more robust schemes that employ limiters or other modifications of boundary fluxes, similar in concept to the limiters employed in finite-volume methods. In contrast to fluid equations, maintaining positivity and conservation simultaneously for kinetic equations can be subtle as the conservation is indirect, involving integration by parts and exchange of energy between particles and fields. We will present a novel algorithm for positivity that continues to maintain total conservation for kinetic systems in a future publication. Finally, DG schemes can also be extended to handle complex geometries that are required to study many laboratory plasmas and industrial applications.

The rest of the paper is organized as follows. We first summarize, in Section~\ref{sec:theory}, properties of the continuous Fokker-Planck operator that are important to maintain or control in the discrete system. Section~\ref{sec:weakeq} describes the concept of weak equality and its use in computing primitive moments and continuous reconstructions. We then present two FPO DG discretizations in Section~\ref{sec:discreteFokkerPlanck} showing why the diffusion operator and velocity space boundary needs to be handled in a particular manner to ensure conservation. Preservation of conservative properties requires special handling of the primitive moments; this is reviewed in Section~\ref{sec:scheme-theory}. Time-integration stability is briefly commented on in Section~\ref{sec:timeStepping}. A series of benchmark problems for both the stand-alone FPO as well as the complete Vlasov-Maxwell-Fokker-Planck system illustrate the physics that can be studied with the scheme in Section~\ref{sec:Results}. We conclude with some highlights of this work and direction for future research.

The schemes are implemented in \gke, an open-source computational plasma physics package. \gke\ contains solvers for Vlasov-Maxwell-Fokker-Planck  equations, gyrokinetic equations~\cite{Shi2017thesis,Shi2017} and multi-fluid moment equations~\cite{Wang:2015kx}. To allow interested readers to reproduce our results, we have made \gke\ input scripts used in this paper available in a public source repository. Instructions to obtain \gke\ and the input files used here are given in the appendix.
\section{The continuous collision operator and its properties}\label{sec:theory}

The form of the Fokker-Planck operator we will consider here will approximate the coefficients appearing in \eqr{\ref{eq:FProsenbluth}}, as\cite{Dougherty1964,Lenard1958}
\begin{align}
    \langle \Delta v_i \rangle_s = -\nu_{ss} (v_i-u_{s,i}) - \sum_{r\ne s} \nu_{sr} (v_i-u_{sr,i}) \label{eq:deltav}
\end{align}
and
\begin{align}
    \langle \Delta v_i \Delta v_j \rangle_s = 2 \nu_{ss}v_{th,s}^2 \delta_{ij} + \sum_{r\ne s} 2 \nu_{sr}v_{th,sr}^2 \delta_{ij}. \label{eq:deltavv}
\end{align}
Here $\nu_{sr}$ is the collision frequency between particles of species $s$ and $r$, and $\v{u}_s$  and $v_{th,s} = \sqrt{T_s/m_s}$ are the mean velocity and  thermal speed of particles of species $s$ respectively and where $T_s$ for non-Maxwellian species is defined below based on the average energy of the particles. The velocities $\v{u}_{sr}$ and the thermal speeds $v_{th,sr} = \sqrt{T_{sr}/m_s}$ are intermediate values that are determined to maintain conservation of total density, momentum and energy. These coefficients are chosen to preserve important properties of the Rosenbluth Fokker-Planck collision operator (which is equivalent to the Landau collision operator), as we discuss below. See below and also Green~\cite{Greene:1973ut} for the analogous problem with the Bhatnagar-Gross-Krook (BGK) operator.

To compute the mean (drift) velocities and thermal speeds that appear in the above equations, we first define the moments
\begin{align}
    M^s_0 &= \langle 1 \rangle_s \label{eq:M0} \\
    \v{M}^s_1 &= \langle \vv \rangle_s \label{eq:M1} \\
    M^s_2 &= \langle v^2 \rangle_s \label{eq:M2}
\end{align}
where the moment operator $\langle \varphi(\vv) \rangle_s$ is defined as
\begin{align}
\langle \varphi(\vv) \rangle_s = \int_{-\infty}^\infty \varphi(\vv) f_s(\v{x},\vv,t) \thinspace d^3\vv. \label{eq:mom-def}
\end{align}
In terms of these moments the mean velocity and thermal speed in three velocity dimensions are determined from $\v{M}^s_{1} = n_s\v{u}_s$ and $M^s_{2} = n_s\v{u}_s^2 + 3 n_s v_{th,s}^2$. For other velocity dimensions replace the 3 with $d_v$, the dimension of the velocity space.

Even though \eqr{\ref{eq:FProsenbluth}} has the form of the Fokker-Planck equation, with the specific choice for the increments \eqr{\ref{eq:deltav}} and \eqr{\ref{eq:deltavv}} this operator is sometimes referred to, in plasma physics, by different names. For example, a linear form of the operator, that did not conserve momentum, was presented by Lenard and Bernstein in~\cite{Lenard1958} to study Landau damping of plasma oscillations in the presence of collisions. The nonlinear and conservative form above was presented first by Dougherty in~\cite{Dougherty1964}. Hence, in the plasma physics literature this operator is commonly referred to as the Lenard-Bernstein operator or the Dougherty operator. In this paper we refer to the conservative operator as the Dougherty-Lenard Fokker-Planck operator (D-FPO). Note that the form of the equation is the same as the generic Fokker-Planck equation. Determining the increments in the operator using Rosenbluth potentials does not change the form of the basic equation and also, as mentioned in the introduction, most of the computational difficulties and solutions developed here carry over to that case.

We next list the properties of the Rosenbluth FPO (or other collision operators) that are important to preserve in a good numerical scheme. The most important of these properties are the conservation of number density of each species, and the total, summed over species, momentum and total energy. The corresponding conservation properties of the collisionless Vlasov-Maxwell system were given in \cite{Juno2018}. The conservation properties are summarized in the following propositions.

\begin{proposition}[Number Density Conservation] \label{prop:densCons}
The FPO conserves number density of each species:
\begin{align}
    \frac{\partial}{\partial t} \langle 1 \rangle_s = 0.
\end{align}
\end{proposition}
\begin{proposition}[Momentum Conservation] \label{prop:momCons}
The FPO conserves total momentum:
\begin{align}
    \frac{\partial}{\partial t} \sum_s \langle m_s \vv \rangle_s = 0
\end{align}
if
\begin{align}
\sum_s \int_{-\infty}^\infty m_s  \langle\Delta{v_i}\rangle_s f_s \thinspace d^3\vv = 0. \label{eq:mom-ct}
\end{align}
\end{proposition}
\begin{proposition}[Energy Conservation] \label{prop:erCons}
The FPO conserves total energy:
\begin{align}
    \frac{\partial}{\partial t} \sum_s  \left\langle \frac{1}{2} m_s \vv^2 \right\rangle_s  = 0 
\end{align}
if
\begin{align}
\sum_s   \int_{-\infty}^\infty m_s \big(
v_i \langle\Delta{v_i}\rangle_s
+ \frac{1}{2} \langle\Delta v_i \Delta v_i \rangle_s
\big) f_s
\thinspace d^3\vv
= 0. \label{eq:en-ct}
\end{align}
\end{proposition}
We remark that for the full VM-FP equation the integration needs to be performed over the full phase-space and not just the velocity space as above,  or a local conservation law can be derived with additional terms involving the divergence of spatial fluxes of particle, momentum, and energy. The proofs follow from the form of the FPO, \eqr{\ref{eq:FProsenbluth}}, and integrating the drag terms by parts once and the diffusion terms twice. We also need to assume that $f(t,\v{x},\vv)\rightarrow 0$ sufficiently fast as $v_i \rightarrow \infty$.


A well constructed discrete operator should ensure that number density, total momentum and total energy are conserved in the discrete sense. Hence, the scheme must satisfy discrete constraints analogous to momentum and energy conservation in the continuous case. In addition, the proofs for the conservation in the continuous case assumed an infinite velocity domain, but this needs to be truncated in the numerical scheme presented here. Hence, as we will show in Section \ref{sec:scheme-theory}, discrete conservation require accounting for the boundary conditions in velocity space while ensuring constraints on the fluctuations are met. This is unlike the collisionless Vlasov-Maxwell case, see Proposition~8 in~\cite{Juno2018}, in which the velocity boundaries do not appear in the proof of conservation of energy, except implicitly in the assumption of zero flux of particles through the maximum velocity boundary on the grid.

Now consider the case of a single species for which we can write the Dougherty FPO with our choice of fluctuations as
\begin{align}
    \pfrac{f}{t} = \gvs \cdot \left( \nu (\vv -\v{u})f + \nu v_{th}^2\gvs f \right) \equiv C[f] \label{eq:lbo}
\end{align}
where the species subscript is dropped. The generalization to multiple species Dougerty operator will be presented in a future work. For this system, we can prove the following properties.
\begin{proposition}[H-Theorem] The total entropy of the system is a non-decreasing function, that is
\begin{align}
\frac{d}{dt}\int_{-\infty}^{\infty} -f \ln{f} \thinspace d^3\vv \ge 0.
\end{align}
\begin{proof}
Let $S = -\int_{-\infty}^{\infty} f \ln{f} \thinspace d^3\vv$. Then we have
\begin{align}
\pfrac{S}{t} = -\int_{-\infty}^{\infty} \pfrac{f}{t}(\ln f + 1) \thinspace d^3\vv.
\end{align}
Write the Fokker-Planck operator as
\begin{align}
\pfrac{f}{t} = \gvs\cdot\v{F}
\end{align}
where $\v{F} = (\vv-\v{u})f + v_{th}^2 \gvs f$. We have dropped $\nu$ without any loss of generality. Substitute to get
\begin{align}
\pfrac{S}{t} = \int_{-\infty}^{\infty} \frac{1}{f}\v{F} \cdot \gvs f \thinspace d^3\vv
\end{align}
after integration by parts and assuming that $\v{F} \rightarrow 0$ as $\vv \rightarrow \pm \infty$ faster than the logarithmic singularity from $\ln f$. Substitute $\gvs f = (\v{F}-(\vv-\v{u})f)/v_{th}^2$ to get
\begin{align}
\pfrac{S}{t} = \frac{1}{v_{th}^2}
\int_{-\infty}^{\infty} \left[ \frac{1}{f}\v{F}^2 -(\vv-\v{u}) \cdot \v{F} \right] \thinspace d^3\vv.
\end{align}
Using the definition of $\v{F}$ the second term in this equation becomes
\begin{align}
\int_{-\infty}^{\infty} \left[ (\v{v}^2-2\v{u}\cdot \vv + \v{u}^2) f + v_{th}^2 (\vv-\v{u})\cdot  \gvs f \right] \thinspace d^3\vv
= 0
\end{align}
using integration by parts and the definitions of moments. Hence
\begin{align}
\pfrac{S}{t} = \frac{1}{v_{th}^2}
\int_{-\infty}^{\infty} \frac{1}{f}\v{F}^2 \thinspace d^3\vv 
\ge 0
\end{align}
as long as $f \ge 0$.
\end{proof}
\end{proposition}
Remark that for the full Vlasov-Maxwell-Fokker-Planck equation the integration needs to be performed over the complete phase-space and not just velocity space as done above. The definition of entropy and the fact that it increases monotonically can be used to prove the following.
\begin{proposition}[Maximum entropy solution] \label{prop:maxEntropySol}
The maximum entropy solution to the Dougherty (and Rosenbluth-Landau) Fokker-Planck operator is the Maxwellian
given by
\begin{align}
    f_M(n,\v{u},v_{th}) = \frac{n}{(2\pi v_{th}^2)^{3/2}}\exp\left(-\frac{(\vv-\v{u})^2}{2v_{th}^2}\right). \label{eq:maxwellian}
\end{align}
\begin{proof}
We wish to maximize the entropy subject to the constraint that density, momentum and energy do not change during the evolution as required by Propositions\thinspace(\ref{prop:densCons})--(\ref{prop:erCons}). Hence we need to find the extremum of
\begin{align}
    S = -\int_{-\infty}^{\infty} f \ln{f} \thinspace d^3\vv
    + \lambda_0 \left(\int_{-\infty}^{\infty} f \thinspace d^3\vv - M_0\right)
    + \gvec{\lambda}_1 \cdot \left(\int_{-\infty}^{\infty} \vv f \thinspace d^3\vv - \v{M}_1 \right)
    + \lambda_2 \left(\int_{-\infty}^{\infty} \vv^2 f \thinspace d^3\vv - M_2\right)
\end{align}
where $\lambda_0, \gvec{\lambda}_1$ and $\lambda_2$ are Lagrange multipliers. Varying this Lagrangian and applying the constraints to determine the Lagrange multipliers, leads to the Maxwellian. Also, as the entropy is monotonically increasing the Maxwellian maximizes the entropy.
\end{proof}
\end{proposition}

\begin{proposition}[Self-adjoint property] The Dougherty Fokker-Planck operator is self-adjoint: for arbitrary functions $g(t,\v{x},\vv)$, $f(t,\v{x},\vv)$ 
\begin{align}
\langle g C[f] \rangle = \langle f C[g] \rangle \label{eq:lbo-adjoint}
\end{align}
with the inner product defined as
\begin{align}
\langle f g \rangle = 
\int_{-\infty}^\infty \frac{1}{f_M}f g \thinspace d^3\vv
\end{align}
where $f_M$ is the Maxwellian that satisfies $C[f_M]=0$.
\begin{proof}
Integrating \eqr{\ref{eq:lbo-adjoint}} by parts we get
\begin{align}
    \langle g C[f] \rangle =
    -\int_{-\infty}^\infty
    \gvs\left(\frac{g}{f_M}\right)
    \cdot
    \left[
      (\vv-\v{u}) f + v_{th}^2 \gvs f
    \right]
    \thinspace d^3\vv.
\end{align}
We have the identity
\begin{align}
v_{th}^2 f_M \gvs\left(\frac{f}{f_M}\right) = (\vv-\v{u}) f + v_{th}^2 \gvs f.
\end{align}
Using this leads to
\begin{align}
      \langle g C[f] \rangle =
    - v_{th}^2 \int_{-\infty}^\infty  
    f_M \gvs\left(\frac{g}{f_M}\right) \cdot \gvs\left(\frac{f}{f_M}\right)
    \thinspace d^3\vv
    \label{eq:selfad}
\end{align}
This is symmetric in $f$ and $g$ from which self-adjoint property follows.
\end{proof}
\end{proposition}
We note that the self-adjointness of the full Rosenbluth-Landau FPO can only be proved for the \textit{linearized} operator. The self adjoint property indicates that the eigenvalues of the operator are all real and hence all solutions are damped. One can show that the eigenfunctions of the operator \eqr{\ref{eq:lbo}} are simply the multi-dimensional tensor Hermite functions~\cite{HarrisBook,Patarroyo:2019uv,1967PhFl...10.1356G,1993PPCF...35..973H} and each mode is damped proportional to the mode number.

If we set $g=f$ in \eqr{\ref{eq:selfad}} we get
\begin{align}
\int_{-\infty}^\infty
\frac{f}{f_M} \pfrac{f}{t}
\thinspace d^3\vv
=
\frac{d}{dt} \int_{-\infty}^\infty \frac{f^2}{f_M} \thinspace d^3\vv
=
      \langle f C[f] \rangle =
    -v_{th}^2 \int_{-\infty}^\infty f_M  \gvs\left(\frac{f}{f_M}\right) \cdot \gvs\left(\frac{f}{f_M}\right) \thinspace d^3\vv
    \le 0.
\end{align}
This shows that the FP operator will decay $f^2/f_M$ integrated over velocity space.

The proof of self-adjoint property shows that $1/f_M$ is the natural weight that must be used in defining an inner-product for the FP equation. In fact, the $L_2$ norm (without the weight) of the distribution is not monotonic in general. To see this, use the definition of the Fokker-Planck operator, and integrate by parts:
\begin{align}
    \frac{d}{dt}\int_{-\infty}^\infty \frac{1}{2} f^2 \thinspace d^3\vv = 
    -\int_{-\infty}^\infty \gvs f\cdot \left((\vv-\v{u})f + v_{th}^2 \gvs f \right) \thinspace d^3\vv
\end{align}
Write the first term as
\begin{align}
    \gvs f\cdot (\vv-\v{u}) f = \gvs\bigg( \frac{1}{2}f^2 \bigg) \cdot (\vv-\v{u}) = \vv \cdot \gvs\bigg( \frac{1}{2}f^2 \bigg) - \gvs\cdot\bigg( \v{u}\frac{1}{2}f^2 \bigg).
\end{align}
The second term is a total derivative and will vanish on integration. This leaves
\begin{align}
    \frac{d}{dt}\int_{-\infty}^\infty \frac{1}{2} f^2 \thinspace d^3\vv = 
    -\int_{-\infty}^\infty \vv \cdot \gvs\bigg( \frac{1}{2}f^2 \bigg)
    + v_{th}^2 |\gvs f|^2 \thinspace d^3\vv.
\end{align}
Performing integration by parts on the first term
\begin{align}
    \frac{d}{dt}\int_{-\infty}^\infty \frac{1}{2} f^2 \thinspace d^3\vv = 
    \int_{-\infty}^\infty \frac{3}{2}f^2 - v_{th}^2 |\gvs f|^2 \thinspace d^3\vv.
\end{align}
For a Maxwellian the right-hand side vanishes but one can construct perturbations on it that may change the sign. To see this, perform a perturbation around a Maxwellian $f = f_M + \delta f$ to get the variation. Without loss of generality, the 1X1V, no-drift case gives,
\begin{align}
    \delta \frac{d}{dt} \int_{-\infty}^\infty \frac{1}{2} f^2 \thinspace dv
    =
    \int_{-\infty}^\infty \left( f_M + 2 v_{th}^2 \frac{\partial^2 f_M}{\partial v^2}\right) \delta f \thinspace dv    
    =
    \int_{-\infty}^\infty \left(-1+\frac{2 v^2}{v_{th}^2}\right)f_M \delta f \thinspace dv.
\end{align}
Clearly, $\delta f$ can be of any sign. This result shows that the $L_2$ norm is not monotonic and the Maxwellian is not the extremum of the $L_2$ norm. Physically, as the drag velocity $\vv-\v{u}$ is compressible the contribution from the drag term can't be turned into a total derivative. The compressibility of the drag term is in contrast to the collisionless case, in which the phase-space velocity is incompressible and hence the phase-space integrated $f^2$ is constant. However, in kinetic theory the essential ingredient is an H-theorem that shows the maximum entropy solution is the Maxwellian, and the Fokker-Planck operator, even with our approximations, possesses one.

Another way to think about the $1/f_M$ weighting of the inner product is to note that it naturally arises when measuring how much a distribution function deviates away from a Maxwellian in terms of entropy.
In other words, writing $f=f_M + \delta f$, then the entropy $S[f] = - \int dv f \ln(f)$ as a functional of $f$ can be written as $S[f_M+\delta f] = S[f_M] - (1/2) \int dv (\delta f)^2 / f_M + \ldots$ through second order.
This expansion is consistent with the result that any deviation away from a Maxwellian is a state of lower entropy.
Note that, to derive this, we have made use of $\int dv v^p\delta f = 0$ for $p=0, 1, 2$ because the Maxwellian $f_M$ has the same 0th through 2nd moments as $f$.
This norm for $\delta f$ is equivalent to a norm on the total $f$, plus a constant, since $\int dv f^2 / f_M = \int dv (f_M + \delta f)^2 / f_M = n_0 + \int dv (\delta f)^2 / f_M$, where the density $n_0 = \int dv f$ is conserved by the collision operator.
This result that $S[f_M+\delta f] = {\rm constant} - (1/2) \int dv f^2 / f_M + \ldots$ shows a relationship between the collision operator causing the entropy to be never decreasing and the $1/f_M$-weighted norm to be never increasing.
\section{Weak equality, weak operators and recovery scheme for diffusion}\label{sec:weakeq}

Before describing the discontiuous Galerkin (DG) scheme for the FPO we discuss the concept of \textit{weak equality, weak operators} and the recovery method for computing second derivatives. These are essential to the construction of our scheme and ensuring that it is conservative and high order accurate. The general idea of weak-equality and weak-operators is also central to other aspects of our algorithm development projects, including ensuring positivity, designing alias-free kinetic and fluid algorithms, and converting DG data to other basis representations. However, here we will only focus on what is needed for discretizing the FPO, leaving the complete discussion of these concepts to another publication~\cite{HakimWeak:2019}.

Consider some interval $I$ and some function space $\mathcal{P}$ spanned by basis set $\psi_k$, $k=1,\ldots,N$. We will define two functions $f$ and $g$ to be \textit{weakly equal} if
\begin{align}
    \int_I (f-g) \psi_k \thinspace dx = 0, \quad \forall k=1,\ldots,N.
\end{align}
We will denote weakly equal functions by $f \doteq g$. Unlikely strong equality, in which functions agree at all points in the interval, weak equality only assures us that the projection of the function on a chosen basis set is the same. However, the function themselves may be quite different than each other with respect to their behaviour, e.g, the function's positivity or monotonicity in the interval. The key strength of this concept is that it allows replacing a function by its weakly equal counterpart in certain terms of the algorithm without changing those terms, e.g., volume integrals in DG, but allowing more desirable properties in other terms, e.g., the surface integrals in DG.

An application of this concept is to weak-operators. To illustrate, say we know the density and momentum at some physical location; see \eqr{\ref{eq:M0}}--\eqr{\ref{eq:M2}}. To determine the velocity we have the relation $M_0 u = M_1$. At first sight one may think that it is trivial to compute the velocity from this expression: $u = M_1/M_0$. However, in a DG scheme we only know the projections of $M_0$ and $M_1$ on the DG function space and not the function itself. Even in a nodal DG scheme a naive division of nodal values is not correct and will lead to aliasing errors (except for $p=1$ polynomials, where one can do division at Gaussian quadrature nodes, though even there limits need to be applied, see below). Hence, more correctly, to find $u$ we need to invert the \textit{weak-operator} equation
\begin{align}
    M_0 u \doteq M_1.
\end{align}
Using the definition of weak-equality above this expression means
\begin{align}
    \int_I (M_0 u - M_1) \psi_k \thinspace dx = 0.
\end{align}
This only determines $u$ to its projection on the function space. Assuming $u$ belongs to the function space we can write $u = \sum_m u_m \psi_k$ leading to the linear system of equations
\begin{align}
\sum_m u_m \int_I M_0 \psi_k \psi_m \thinspace dx = \int_I M_1 \psi_k \thinspace dx.
\end{align}
for $k=1,\ldots,N$. Inverting this determines $u_m$ and hence the projection of $u$ in the function space. We call this process \textit{weak-division}. Note that weak-division only determines $u$ to an equivalence class as we can replace the specific $u$ in the function space with any other function that is weakly equal to it. In addition, note that $M_0$ and $M_1$ themselves have expansions that must be included in this computation. Weak-division is extensively used in our algorithm to determine ``primitive'' moments like $\v{u}_s$ and $v_{th,s}$ required in computing the fluctuations.

\begin{figure}
  \setkeys{Gin}{width=0.65\linewidth,keepaspectratio}
  \incfig{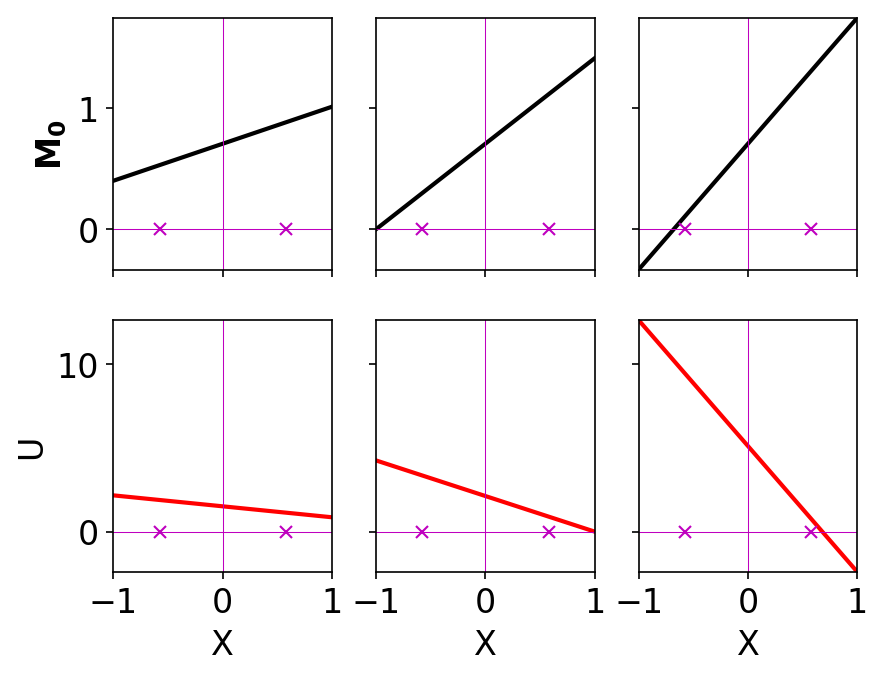}%
   \caption{Weak division for $p=1$ basis to compute $u$ from $M_0 u \doteq M_1$. In this plot $M_1 = 1$ and the effect of changing $M_0$ (top row) on speed (bottom row) is shown. As the density steepens, the velocity becomes larger and blows up when the density has a zero crossing at $x = \pm 1/\sqrt{3}$ (magenta crosses). To avoid problems, we set a limit on the slope of $M_0$ used in calculating $u(x)$, which still conserves the cell averaged momentum.}
   \label{fig:weak-div-p1}
\end{figure}

One can't divide by zero in regular division. An analogous situation exists for weak-division. To illustrate, consider the interval $[-1,1]$ and the orthonormal linear basis set
\begin{align}
    \psi_0 = \frac{1}{\sqrt{2}}; \qquad \psi_1 = \frac{\sqrt{3}}{\sqrt{2}}x.
\end{align}
Let $M_1 = 1$ and $M_0 = n_0\psi_0 + n_1 \psi_1$. For this simple case the result of weak-division is 
\begin{align}
    u = \frac{\sqrt{2}}{n_0^2-n_1^2}(n_0 - \sqrt{3}n_1 x)
\end{align}
Hence, the weak-division is not defined for $n_1 = \pm n_0$ ($n_0>0$ as mean density must be positive). This shows that, even if the mean density is positive, the slope cannot become too steep. When the ``blow-up'' occurs, i.e., $n_1 = \pm n_0$ case, $M_0$ has a zero-crossing at either $x = \pm 1/\sqrt{3}$. (Note that there is nothing necessarily unphysical with a piecewise linear reconstruction $M_0(x)$ having a zero crossing within the domain (and the DG algorithm can result in such solutions), since in principle there is a physically realizable function that is weakly equivalent $\tilde{M}_0(x) \doteq M_0(x)$ but positive everywhere, as long as $n_1 < \sqrt{3} n_0$). Hence, one must use some form of limiters to ensure that the density does not become too steep. In practice, we have found that the bound $|n_1| < n_0$ is not sufficiently restrictive, as it still allows $u(x)$ to approach infinity, giving rise to severe CFL limits on the time step.  Instead, we restrict $|n_1| < n_0 / \sqrt{3}$ in the calculation of $u(x)$.  Even if we simply set $n_1=0$ and so calculated the mean velocity $u(x)$ by dividing the momentum $M_1(x)$ by the mean density, we would still conserve momentum averaged over a cell.  There is some numerical diffusion of momentum within a cell caused by applying a limiter to $n_1$, but no numerical diffusion across cell boundaries.  (Note that this is similar to the philosophy of limiters in high order finite-volume methods.  In smooth regions we use the standard calculations and so retain high-order accuracy there, while introducing limiters where the solution is locally varying too quickly to be accurately resolved, in order to robustly preserve certain properties of the solution.). Further discussion will be presented in~\cite{HakimWeak:2019}.

Another application of weak-equality is to recover a continuous function from a discontinuous one. Say we want to construct a continuous representation $\hat{f}$ on the interval $I=[-1,1]$, from a function, $f$, which has a single discontinuity at $x=0$. We can choose some function spaces $\mathcal{P}_L$ and $\mathcal{P}_R$ on the interval $I_L = [-1,0]$ and $I_R = [0,1]$ respectively. Then, we can reconstruct a continuous function $\hat{f}$ such that
\begin{align}
  \hat{f} &\doteq f_L \quad x \in I_L \quad\mathrm{on}\
      \mathcal{P}_L \label{eq:he1} \\
  \hat{f} &\doteq f_R \quad x \in I_R \quad\mathrm{on}\ \mathcal{P}_R. \label{eq:he2}
\end{align}
where $f = f_L$ for $x\in I_L$ and $f = f_R$ for $x\in I_R$. Again, this only determines $\hat{f}$ to its projections in the left and right intervals. To determine $\hat{f}$ uniquely, we use the fact that given the $2N$ pieces of information, where $N$ is the number of basis functions in $\mathcal{P}_{L,R}$, we can construct a polynomial of maximum order $2N-1$. We can hence write
\begin{align}
  \hat{f}(x) = \sum_{m=0}^{2N-1} \hat{f}_m x^m.
\end{align}
Using this in Eqn.\thinspace (\ref{eq:he1}) and (\ref{eq:he2}) completely determines $\hat{f}$. In a certain sense the recovery procedure is a special case of a more general method to go from one basis to another under the restriction of weak equality. This allows, for example, to extract Fourier and Hermite moments from DG simulation data. These, and other applications of weak equality and weak operators, will be presented in~\cite{HakimWeak:2019}.

Given this procedure to recover a continuous function, we can now compute second derivatives as follows. We wish to compute $g \doteq f_{xx}$ where we know $f$ on a mesh with cells $I_j = [x_{j-1/2}, x_{j+1/2}]$. Multiply by some test function $\varphi \in \mathcal{P}_j$, where $\mathcal{P}_j$ the function space in cell $I_j$ and integrate to get the following weak-form,
\begin{align}
    \int_{I_j} \varphi g \thinspace dx =   \varphi \hat{f}_x \bigg|^{x_{j+1/2}}_{x_{j-1/2}}
  -
  \int_{I_j} \varphi_{x} f_x \thinspace dx. \label{eq:gfxxp1}
\end{align}
Where we have replaced $f$ by the reconstructed function $\hat{f}$ in the surface term. Note that we need \textit{two reconstructions}, one using data in cells $I_{j-1}, I_j$ and the other using data in cells $I_j, I_{j+1}$. In the volume term we continue to use $f$ itself and not the left/right reconstructions as the latter are weakly-equal to the former and can be replaced without changing the volume term. Once the function space is selected this completely determines $g$. 

Notice that one more integration by parts can be performed in \eqr{\ref{eq:gfxxp1}} to obtain another weak-form,
\begin{align}
    \int_{I_j} \varphi g \thinspace dx =   (\varphi \hat{f}_x - \varphi_x \hat{f})\bigg|^{x_{j+1/2}}_{x_{j-1/2}}
  +
  \int_{I_j} \varphi_{xx} f \thinspace dx. \label{eq:gfxxp2}
\end{align}
In this form we need to use both the value and first derivative of the reconstructed functions at the cell interfaces. Numerically, each of these weak-forms will lead to different update formulas. For example, for piecewise linear basis functions the volume term drops out in \eqr{\ref{eq:gfxxp2}}. Also, the properties of each scheme will be different. In fact, as we show in Sections \ref{sec:discreteFokkerPlanck} and \ref{sec:scheme-theory}, the second integration by parts is required to ensure momentum and energy conservation. (One way to think of this is that \eqr{\ref{eq:gfxxp1}} is not uniquely defined if $f$ is discontinuous at cell boundaries, in which case $f_x$ contains delta functions at the interval boundaries and a consistent prescription for how to handle that has not yet been defined. Equation\thinspace\ref{eq:gfxxp2} provides such a prescription. If one simply ignores the delta functions in \eqr{\ref{eq:gfxxp1}}, then significant errors can occur.)

The procedure outlined above is essentially the recovery discontinuous Galerkin (RDG) scheme first proposed by van Leer in~\cite{VanLeer2005,VanLeer2007}. Extensive study of the properties of the RDG scheme to compute second derivatives is presented in~\cite{Hakim2014} where it is shown that the RDG scheme has some advantages compared to the standard local discontinuous Galerkin (LDG) schemes~\cite{Cockburn1998} traditionally used to discretize diffusion operators in DG. The formulation in terms of weak equality allows systematic extension to higher dimensions and cross-derivative terms, both of which are needed to discretize the Fokker-Planck operator.
\section{The discrete Dougherty Fokker-Planck operator}\label{sec:discreteFokkerPlanck}

Although the FPO was formulated rather generally in Section~\ref{sec:theory}, in order to give a clear explanation of the computational method we will focus, without loss of generality, on the 1X1V case, that is with one configuration space and one velocity space dimensions. The cross-species collision terms are set to zero. This simplified setting is enough to describe the method and the key ideas needed to construct conservative schemes. Some notes on the multi-dimensional and cross-species collisions cases are given at the end of this section. We have implemented the FPO in multiple dimensions in \gke, and we show multi-dimensional results in Section \ref{sec:Results}. 

In the 1X1V case the Dougherty FPO is,
\begin{align}
    \pfrac{f}{t} = \frac{\partial}{\partial v}
\left[
  (v-u) f + v_{th}^2 \pfrac{f}{v}
\right].
\end{align}
Here, we are setting $\nu=1$ (or effectively rescaling the time-coordinate by the collision frequency), and note that the drift velocity and thermal speeds are, in general, spatially dependent quantities. We use a discontinuous Galerkin (DG) scheme on a rectangular mesh with cells $\Omega_{i,j} \equiv [x_{i-1/2},x_{i+1/2}]\times [v_{j-1/2},v_{j+1/2}]$ and select a set of orthonormal basis functions on each cell $\psi_k(x,v)$, for $k=1,\ldots,N$:
\begin{align}
    \int_{\Omega_{i,j}} \psi_k \psi_m \dxdv = \delta_{km}.
\end{align}
To construct these basis functions, first a set of monomials is selected and then orthonormalized using the Gram-Schmidt procedure. The monomials are constructed to lie in the Serendipity polynomial space~\cite{Arnold:2011eu} of order $p$ and dimension $d$, $\mathcal{V}_d^p$, and are constructed, in 2D, as follows
\begin{align}
    \mathcal{V}_2^p = \{ x^m v^n \mid \mathrm{deg}_2(x^m v^n) \leq p \}.
\end{align}
Where the $\mathrm{deg}_2$ function is defined as the sum of all monomial powers that appear superlinearly. For example, $\mathrm{deg}_2(x v^2) = 2$ while $\mathrm{deg}_2(x^2 v^2) = 4$. This definition works in arbitrary dimensions as shown in~\cite{Arnold:2011eu}. For higher dimensional systems and for $p>1$ the reduction in the number of degrees-of-freedom (DOF) compared to Largange-tensor polynomials can be significant. For example, in 5D $p=2$ case there are $3^5 = 243$ Lagrange tensor basis functions while there are only $112$ Serendipity basis functions.  In general, the cost of the modal DG scheme presented here scales roughly as $N^2$ and so Serendipity basis can give $4\times$ cost savings. One could use an even more sparse basis-set, for example by using $\deg$ function that sums all the monomial powers. In this space, $\mathrm{deg}(x v^2) = 3$. With this, in 5D and $p=2$, one would only have 21 basis functions, but we have found that the Serendipity basis a good compromise between efficiency and accuracy. The algorithms presented below do not rely on the specific choice of the basis-sets in any case.

We remark that the configuration space quantities like drift velocity, thermal speed and moments lie in the space $\mathcal{V}^p_1$ (for the 1X1V case). The discrete moments are computed using the weak-equality relations
\begin{align}
    M_{h,0} &\doteq \int f_h \thinspace dv \label{eq:Mh0} \\
    M_{h,1} &\doteq \int v f_h \thinspace dv \label{eq:Mh1} \\
    M_{h,2} &\doteq \int v^2 f_h \thinspace dv \label{eq:Mh2}
\end{align}
on the space $\mathcal{V}^p_1$. The integrals are taken over all velocity space. The generalization of the function spaces and moment definitions to arbitrary velocity- and configuration-space dimensions is straightforward.

A first version of the scheme can be constructed by multiplying the FPO by a test function $w\in\mathcal{V}_2^p$ and, as would be natural for a DG scheme, integrating by parts once. That is we need to find $f_h \in \mathcal{V}_2^p$ such that
\begin{align}
\int_{\Omega_{i,j}} w \pfrac{f_h}{t} \dxdv
=
-\int_{x_{i-1/2}}^{x_{i+1/2}} w
G(f_L,f_R) dx \Bigg|_{v_{j-1/2}}^{v_{j+1/2}} 
-
\int_{\Omega_{i,j}} \pfrac{w}{v}
\left[ (v-u)f_h + v_{th}^2 \pfrac{f_h}{v}
\right] \dxdv.    \label{eq:schemenc}
\end{align}
Here, $G(f_L,f_R)$ is the numerical flux written in a \emph{penalty} form
\begin{align}
G(f_L,f_R) = -\frac{1}{2}(v-u)(f_R+f_L) + \frac{\tau}{2}(f_L-f_R) - v_{th}^2 \pfrac{\hat{f}_h}{v}
\end{align}
where $\tau = \max(|u-v|)$, the maximum computed over the velocity domain at each configuration space cell. The quantity $\hat{f}_h$ appearing in the numerical flux is the \textit{recovered distribution function} and is computed using the weak-equality procedure outlined in Section \ref{sec:weakeq}. See below on how to determine the discrete drift velocity and thermal speed. The scheme \eqr{\ref{eq:schemenc}} is called {\bf Scheme NC}. NC stands for ``non-conservative'' as is shown below.

The velocity space in the FP equation has $v \in [-\infty,\infty]$. In a numerical scheme one must somehow treat this infinite domain in a tractable manner. There are two possible options. First, to assume a form of the distribution function beyond some $v<v_{min}$ ($v>v_{max}$), and update these ``half-open'' cells as part of the scheme. The second approach, followed here, is to simply truncate the domain and assume that instead $v \in [v_{min},v_{max}]$ for some sufficiently small $v_{min}$ and sufficiently large $v_{max}$. In this case one must apply boundary conditions by specifying the numerical flux at $v_{min}$ and $v_{max}$. To ensure that particles are not ``lost'' to infinity we chose the boundary condition
\begin{align}
    G\big(f_L(v_{\min}),f_R(v_{\min})\big) = 
    G\big(f_L(v_{\max}),f_R(v_{\max})\big)
    = 0 \label{eq:fluxbc}
\end{align}
This insures that one does not exchange density, momentum or energy with ``infinity''.

Although Scheme NC seems perfectly reasonable (and would be the natural thing to do based on experience in discretizing hyperbolic equations), in fact, the scheme does not conserve momentum or energy. Density conservation follows on using $w=1$ as the test function and summing over velocity space and using the boundary conditions, \eqr{\ref{eq:fluxbc}}. However, momentum or energy are not conserved. To show the lack of momentum conservation, we choose $w = v$ in \eqr{\ref{eq:schemenc}} and sum over velocity space to obtain,
\begin{align}
\sum_j 
\int_{\Omega_{i,j}} v \pfrac{f_h}{t} \dxdv
=
-
\sum_j 
\int_{\Omega_{i,j}} \left[ (v-u)f_h + v_{th}^2 \pfrac{f_h}{v} \right] \dxdv,
\end{align}
where the continuity of the numerical flux and boundary conditions were used to show the contribution to momentum from surface terms is zero. The first term on the right-hand vanishes if the discrete moments are computed to satisfy
\begin{align}
    \int_{x_{i-1/2}}^{x_{i+1/2}} M_{h,1} - u M_{h,0} \thinspace dx = 0,
\end{align}
where $M_{h,0}$ and $M_{h,1}$ are the discrete moment operators corresponding to the continuous operators, see \eqsr{\ref{eq:Mh0}}{\ref{eq:Mh1}}. This condition is discussed in more detail when presenting the conservative scheme in Section \ref{sec:scheme-theory}. The second term, on integration by parts becomes
\begin{align}
    -\sum_j \int_{x_{i-1/2}}^{x_{i+1/2}} v_{th}^2 f_h \thinspace dx \Bigg|_{v_{j-1/2}}^{v_{j+1/2}},
\end{align}
where $f_h$ must be evaluated just inside the cell. However, in general, $f_h$ is not continuous so this term will not vanish, and instead pick up the sum of jumps in the distribution function across the velocity direction. Hence, \textit{Scheme NC does not conserve momentum}. A similar argument with $w=v^2$ shows that \textit{Scheme NC does not conserve energy either}.

The lack of conservation comes about from the gradient in the volume term. This suggests that instead \textit{two integration} by parts should be performed to remove this gradient. That is for $f_h \in \mathcal{V}_2^p$ the weak-form describing the second scheme is,
\begin{align}
\int_{\Omega_{i,j}} w \pfrac{f_h}{t} \thinspace dx\thinspace dv
=
-\int_{x_{i-1/2}}^{x_{i+1/2}} w
G(f_L,f_R) dx \Bigg|_{v_{j-1/2}}^{v_{j+1/2}} 
- 
\int_{x_{i-1/2}}^{x_{i+1/2}} \pfrac{w}{v} v_{th}^2 \hat{f}_h
 dx \Bigg|_{v_{j-1/2}}^{v_{j+1/2}} 
-
\int_{\Omega_{i,j}} 
\left[
\pfrac{w}{v} (v-u)f_h - \frac{\partial ^2 w}{\partial v^2} v_{th}^2 f_h
\right]
\thinspace dx\thinspace dv, \label{eq:schemec}
\end{align}
for all $w \in \mathcal{V}_2^p$. For this scheme, we need to use the recovered distribution function $\hat{f}_h$ and its derivative at cell boundaries. This scheme is called {\bf Scheme C} and will be the focus of rest of the paper.
\section{The properties of the discrete Dougherty Fokker-Planck operator. Drift-velocity and thermal-speed} \label{sec:scheme-theory}

Having presented the discrete scheme and the motivation for the second integration by parts, we now study its properties. In particular, we examine the scheme's conservation properties. As we show below the Scheme C described by \eqr{\ref{eq:schemec}} conserves particles as well as momentum and energy. Interestingly, the conditions for conservation also provide recipes to determine the drift-velocity and thermal speeds, thus completing the description of the scheme. 

\begin{proposition}[Discrete Number Density Conservation] \label{prop:sc_densCons}
Scheme C conserves number density:
\begin{align}
    \frac{d}{d t} \sum_{j}  \int_{\Omega_{ij}} f_h \dxdv = 0.
\end{align}
\begin{proof}
To prove this, use $w=1$ in \eqr{\ref{eq:schemec}} and sum over all velocity space cells. Using \eqr{\ref{eq:fluxbc}} for the boundary conditions at $v_{min}$ and $v_{max}$, and noting that the numerical flux is continuous at cell interfaces in the interior of the domain, makes the surface integrals in the sum over all velocity space a telescopic sum, completing the proof.
\end{proof}
\end{proposition}
We remark that the sum above, and in all the propositions in this section, is over velocity space cells as there are no spatial gradients in the FP equation. For the full Vlasov-Maxwell Fokker-Planck equation, however the sum extends over all of phase-space.

\begin{proposition}[Discrete Momentum Conservation] \label{prop:sc_momCons}
Scheme C conserves momentum:
\begin{align}
    \frac{d}{d t} \sum_{j}  \int_{\Omega_{ij}} v f_h \dxdv = 0.
\end{align}
if the following \textit{weak-equality relation} is satisfied:
\begin{align}
    v_{th}^2 \big( f_h(v_{max})-f_h(v_{min})\big) + M_{h,1} - u M_{h,0} \doteq 0. \label{eq:momWeakConst}
\end{align}
\begin{proof}
To prove this, use $w=v$ in \eqr{\ref{eq:schemec}} and sum over all velocity space cells to get
\begin{align}
    \frac{d}{d t} \sum_{j}  \int_{\Omega_{ij}} v f_h \dxdv
    =
    -
     \sum_j \int_{x_{i-1/2}}^{x_{i+1/2}} v_{th}^2 \hat{f}_h \thinspace dx \Bigg|_{v_{j-1/2}}^{v_{j+1/2}}
    -
    \sum_j
\int_{\Omega_{i,j}} (v-u)f_h \thinspace dx\thinspace dv.
\end{align}
The contributions from the numerical flux $G$ drop out due to continuity and boundary conditions. In the first term all interface contributions except the first and last will cancel. Combined with the definition of the discrete moment operators in the second term leads to the constraint
\begin{align}
    \int_{x_{i-1/2}}^{x_{i+1/2}}
    \left[
    v_{th}^2\big(f_h(v_{max}) - f_h(v_{min}) \big)
    +
    M_{h,1} - u M_{h,0}
    \right] \thinspace dx
    = 0. \label{eq:momConst}
\end{align}
Using the definition of weak-equality this implies that the momentum will be conserved if \eqr{\ref{eq:momWeakConst}} is satisfied. Notice that the $\hat{f}_h$ was replaced by $f_h$. At the outer velocity boundaries there is no ``outside'' cell to allow reconstructing a distribution function. 
\end{proof}
\end{proposition}
The weak-equality constraint, \eqr{\ref{eq:momWeakConst}}, is stronger than required to satisfy \eqr{\ref{eq:momConst}}. However, ensuring that the weak-equality constraint is satisfied automatically ensures that the conditions for momentum conservation are met.

Remark that the first term in \eqr{\ref{eq:momWeakConst}} is generally a small correction for a well resolved simulation where $f_h$ at the velocity boundaries is small, but it is retained to give exact momentum conservation to within roundoff error.  The origin of this term can be understood by looking at momentum conservation with the continuous operator on a finite velocity domain. 

\begin{proposition}[Discrete Energy Conservation] \label{prop:sc_erCons}
Scheme C conserves energy:
\begin{align}
    \frac{d}{d t} \sum_{j}  \int_{\Omega_{ij}} \frac{1}{2} v^2 f_h \dxdv = 0.
\end{align}
if the following weak-equality relation is satisfied:
\begin{align}
    v_{th}^2 \big( v_{max}f_h(v_{max})-v_{min} f_h(v_{min})\big) + M_{h,2} - u M_{h,1} -v_{th}^2 M_{h,0} \doteq 0. \label{eq:erWeakConst}
\end{align}
\begin{proof}
For this proposition assume that $p\ge 2$ and so $v^2 \in \mathcal{V}_2^p$. The case of $p=1$ introduces some complexity, dealt with below. To prove the proposition use $w=v^2/2$ in \eqr{\ref{eq:schemec}} and sum over all velocity space cells to obtain,
\begin{align}
    \frac{d}{d t} \sum_{j}  \int_{\Omega_{ij}} \frac{1}{2} v^2 f_h \dxdv
    =
    -
     \sum_j \int_{x_{i-1/2}}^{x_{i+1/2}} v v_{th}^2 \hat{f}_h \thinspace dx \Bigg|_{v_{j-1/2}}^{v_{j+1/2}}
    -
    \sum_j
\int_{\Omega_{i,j}} \left[ 
v (v-u)f_h - v_{th}^2 f_h
\right]
\dxdv.
\end{align}
All contributions from interior cell interfaces cancel in the first term. Combined with the definition of the discrete moment operators in the second term leads to the constraint,
\begin{align}
    \int_{x_{i-1/2}}^{x_{i+1/2}}
    \left[
    v_{th}^2\big(v_{max}f_h(v_{max}) - v_{min}f_h(v_{min}) \big)
    +
    M_{h,2} - u M_{h,1} - v_{th}^2 M_{h,0}
    \right] \thinspace dx
    = 0. \label{eq:erConst}
\end{align}
Using the definition of weak-equality this implies that the energy will be conserved if \eqr{\ref{eq:erWeakConst}} is satisfied.
\end{proof}
\end{proposition}

Collecting the constraints needed to satisfy momentum and energy conservation we have the following simultaneous set of weak-equality relations
\begin{align}
    v_{th}^2 \big( f_h(v_{max})-f_h(v_{min})\big) + M_{h,1} - u M_{h,0} &\doteq 0, \\
    v_{th}^2 \big( v_{max}f_h(v_{max})-v_{min} f_h(v_{min})\big) + M_{h,2} - u M_{h,1} -v_{th}^2 M_{h,0} &\doteq 0.
\end{align}
We can express these two equations as a linear system that, on inversion, determines the drift velocity $u$ and the thermal speed $v_{th}$ in $\mathcal{V}_1^p$, completing the description of the scheme. There are some interesting aspects of this derivation. First, the requirements of discrete momentum and energy conservation leads to \textit{unique} expressions that determine the drift velocity and thermal speed. Second, the boundary conditions in velocity space must be accounted for in the computations. In fact, our first impulse, following experience in discretizing the Vlasov equation had us expect that the energy conservation would follow independent of boundary conditions. An unexpected result of carefully working through the conservation proofs shows otherwise.

The energy conservation proof above requires $p\ge 2$. In the the case of $p=1$, instead a projected energy is conserved. To see this first observe that definition of discrete particle energy is
\begin{align}
    E_h = \int_{-\infty}^\infty \frac{1}{2} m v^2 f_h \thinspace d^3\vv.
\end{align}
Using the idea of weak-equality, we see that this integral is unchanged if we replace $v^2$ with another function that is weakly equal to it. For use in energy conservation we will chose the projection on the basis set, that is $\overline{v^2} \in \mathcal{V}_2^1$ defined such that
\begin{align}
    \overline{v^2} \doteq v^2 \quad\mathrm{on}\ \mathcal{V}_2^1. \label{eq:v2bar}
\end{align}
A simple calculation shows that the function $\overline{v^2}$ is continuous. With this definition of projected $v^2$ we can now state the following proposition that leads to an analogous weak-equality relation as \eqr{\ref{eq:erWeakConst}}.
\begin{proposition}[Discrete Energy Conservation, $p=1$ case] \label{prop:sc_erCons_p1}
Scheme C conserves energy:
\begin{align}
    \frac{d}{d t} \sum_{j}  \int_{\Omega_{ij}} v^2 f_h \dxdv = 0.
\end{align}
if the following weak-equality relation is satisfied:
\begin{align}
    v_{th}^2 \big( \vcen_{max}f_h(v_{max})-\vcen_{min} f_h(v_{min})\big) + M^*_{h,2} - u M^*_{h,1} -v_{th}^2 M^*_{h,0} \doteq 0. \label{eq:erWeakConstP1}
\end{align}
where $\vcen_j = (v_{j+1/2}+v_{j-1/2})/2$ is the cell-center velocity coordinate. The ``star moments'' $M^*_{h,k}$ are defined below.
\begin{proof}
As $v^2/2$ does not belong in $\mathcal{V}_2^1$ we use its projection to show energy conservation. That is use $w = \overline{v^2}/2$, see \eqr{\ref{eq:v2bar}}. With this choice we can show that
\begin{align}
    \pfraca{v} \left(\frac{1}{2} {\overline{v^2}} \right) = \frac{1}{2}(v_{j+1/2} + v_{j-1/2}) \equiv \vcen_j.
\end{align}
Setting $w = \overline{v^2}/2$  in \eqr{\ref{eq:schemec}} and summing over all velocity space cells we obtain,
\begin{align}
    \frac{d}{d t} \sum_{j}  \frac{1}{2} \int_{\Omega_{ij}} v^2 f_h \dxdv
    =
     -
     \sum_j \int_{x_{i-1/2}}^{x_{i+1/2}} \vcen_j v_{th}^2 \hat{f}_h \thinspace dx \Bigg|_{v_{j-1/2}}^{v_{j+1/2}}
    -
    \sum_j
\int_{\Omega_{i,j}} \vcen_j (v-u)f_h  \dxdv.
\end{align}
As before, the contribution from the numerical flux term drops out as $\overline{v^2}/2$ is continuous. However, since $\vcen_j$ is not continuous the contributions from the first term above cannot be eliminated. This result can be written as,
\begin{align}
    \int_{x_{i-1/2}}^{x_{i+1/2}} v_{th}^2 \sum_j  \vcen_j  \hat{f}_h \Bigg|_{v_{j-1/2}}^{v_{j+1/2}} \thinspace dx 
    =
    \int_{x_{i-1/2}}^{x_{i+1/2}}
    \left[
    v_{th}^2 \big( \vcen_{max}f_h(v_{max})-\vcen_{min} f_h(v_{min})\big)
    - v_{th}^2 \sum_{j \neq \pm j_{max}} \Delta v_j \hat{f}_{h,{j+1/2}}
    \right]
    \thinspace dx
\end{align}
where $\Delta v_j = \vcen_{j+1}-\vcen_j$ and $j \neq \pm j_{max}$ is taken to mean that the sum excludes the first and last $j$ indices. Defining the ``star moments'' as,
\begin{align}
    M^*_{h,0} &\doteq \sum_{j \neq \pm j_{max}} \Delta v_j \hat{f}_{h,j+1/2}, \\
    M^*_{h,1} &\doteq \sum_j \vcen_j \int_{v_{j-1/2}}^{v_{j+1/2}} f_h \thinspace dv, \\
    M^*_{h,2} &\doteq \sum_j \vcen_j \int_{v_{j-1/2}}^{v_{j+1/2}} v f_h \thinspace dv,
\end{align}
and using the definition of weak-equality, the projected energy will be conserved in the $p=1$ case if \eqr{\ref{eq:erWeakConstP1}} is satisfied.
\end{proof}
\end{proposition}
In summary, for the $p=1$ case the drift velocity and thermal speed must be determined using the following simultaneous set of weak-equality relations
\begin{align}
    v_{th}^2 \big( f_h(v_{max})-f_h(v_{min})\big) + M_{h,1} - u M_{h,0} &\doteq 0 \\
    v_{th}^2 \big( \vcen_{max}f_h(v_{max})-\vcen_{min} f_h(v_{min})\big) + M^*_{h,2} - u M^*_{h,1} -v_{th}^2 M^*_{h,0} &\doteq 0.
\end{align}
In this case, one must compute the discrete star moments and the regular moments (except for $M_{h,2}$).

We remark that we have not managed to analytically show that the discrete operator increases entropy, although, as shown in the benchmarks section, all our simulations increase entropy monotonically. Part of the difficulty is computing $w \doteq \ln f_h$ and its needed gradients. Also, the proof of self-adjoint property (which we have not yet worked out for the general DG algorithm) requires the definition of a ``discrete Maxwellian'' for use in the appropriate inner-product norm. 

Finally, we remark that the scheme and proofs above are for the 1X1V case but the extension to multiple dimensions is straightforward. The extension of the conservative scheme to cross-species collisions is much more involved and requires carefully computing the intermediate velocities and thermal speeds while taking into account finite boundary effects. In addition, there are some issues in ensuring that the intermediate temperatures remain positive (even in the continuous case) and we will discuss these in a future publication.
\section{Time-stepping and stability} \label{sec:timeStepping}

The focus of this paper is on the spatial discretization of the Fokker-Planck operator and showing that a carefully constructed scheme is conservative. The issue of time-stepping is a complex one, specially as the FPO is diffusive and hence for efficient evolution (when collision frequency is high) requires some implicit or accelerated time-stepping scheme. As we are focused on the spatial scheme, for the results presented in the benchmarks sections we use an explicit three-stage Strong-Stability Preserving Runge-Kutta third order scheme~\cite{shu2002} (SSP-RK3). Exploring more complex time-steppers is left to a future study. The time-step is selected such that $\lambda_{\rm max} \Delta t$ is small enough that the solutions are stable, where the maximum eigenvalue of the discretized collision operator scales approximately as $\lambda_{\rm max} \sim \nu( (v_{\rm max}+ |u|)/(\Delta v) + v_{th}^2 / (\Delta v)^2)$. The issue of stability for the full Vlasov-Maxwell Fokker-Planck equation is complex and will be addressed in a paper on discretizing the gyrokinetic version of the operator.

While this works fine in lower collisionality regimes, for efficient simulations at higher collisionality (which can be important in many experiments), one would have to use some sort of implicit or accelerated methods. For example, we have successfully used the Runge-Kutta-Legendre (RKL) ``Super Time-Stepping'' methods~\cite{Meyer:2014hm}. This time-stepper is composed of forward-Euler stages, the number of stages and update formulas determined dynamically based on the ratio of size of time-step desired to explicit time-step allowed. For the case of VM-FP system, of course, the ratio must account for the stability limits of the collisionless terms also, usually updated with an explicit scheme. For an $s$ stage scheme a collisional time-step on the order of $s^2 \Delta t_{explicit}$ can be taken, giving an effective speed-up of $s$. Depending on the regime, $s$ can be large (on the order of a few tens of stages). 
\section{Benchmark problems}\label{sec:Results}

The algorithms presented above are being used in production simulations with the \gke\ code. For example, recent work has explored the role of collisions in electrostatic shocks~\cite{Sundstrom:2018wb}, and much of our current work is focused on studying the effect of weak collisions on space and laboratory plasmas. An earlier version of the scheme was applied to gyrokinetic equations and was used to study plasma turbulence in straight and helical field-line geometries~\cite{Shi2017thesis,Shi2017,Shi2019}, including comparison with experiments\cite{Bernard:2018wb}. In this section we present benchmark problems that allow testing the scheme in simpler settings, in particular, paying attention to the conservation properties and the ability of the scheme to reproduce results known analytically or with other schemes. The tests are arranged in increasing order of complexity. First, we benchmark just the collision operator with spatially homogeneous relaxation problems. Next we introduce streaming terms and study shocks in a neutral gas. We then turn to electrostatic plasma problem and benchmark the scheme with collisional Landau damping. Finally, as an application of the full Vlasov-Maxwell-Fokker-Planck equation we study plasma heating due to ``magnetic pumping'', a process that converts electromagnetic wave energy into heating via pitch-angle scattering.

\subsection{Relaxation to discrete Maxwellian}

In the absence of streaming and body forces, any initial distribution function will quickly relax to a Maxwellian. This is evident as the Fokker-Planck operator increases entropy and the maximum entropy solution is the Maxwellian; see Proposition\thinspace\ref{prop:maxEntropySol}. A subtle point is that on a discrete velocity grid the relaxed distribution function is \textit{by definition} the discrete Maxwellian. One may at first sight be tempted to define a ``discrete Maxwellian'' as the projection of \eqr{\ref{eq:maxwellian}} on to basis functions. However, the derivation that led to \eqr{\ref{eq:maxwellian}} assumed a continuous and infinite velocity domain. Hence, if the derivation is repeated with the specific form of the discrete scheme, the ``discrete Maxwellian'' will not be the same as the projection of \eqr{\ref{eq:maxwellian}}, though they will converge towards each other as the grid is refined. In fact, this relaxation process, rather than projection, can be used to define a discrete Maxwellian for use in initializing other simulations.

\begin{figure}
  \setkeys{Gin}{width=0.45\linewidth,keepaspectratio}
  \incfig{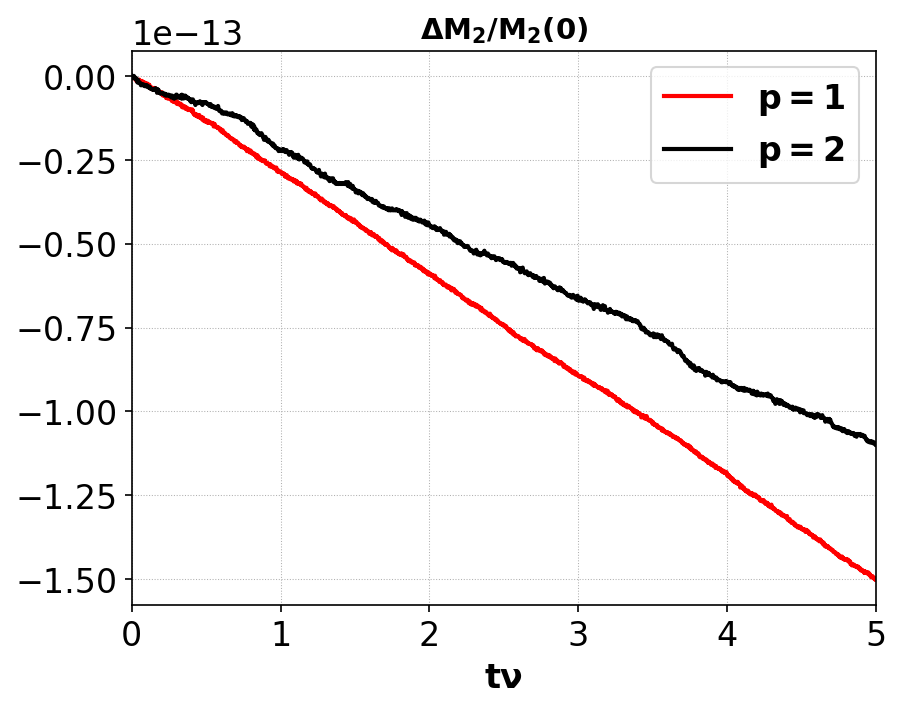}%
   \incfig{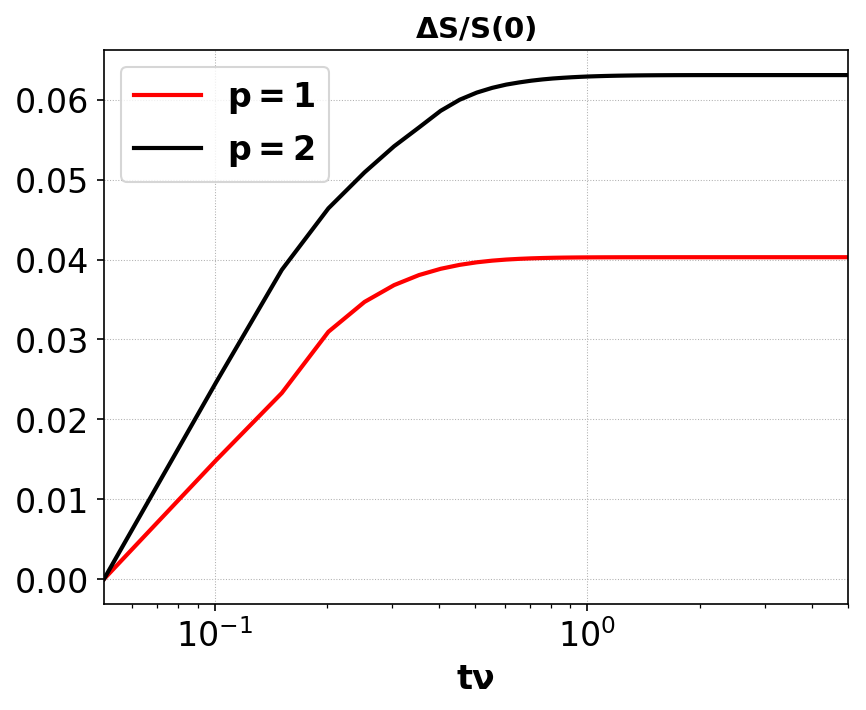}%
   \caption{Relative change in energy (left) for $p=1$, $N=16$ (red) and $p=2$, $N=8$ (black) cases for relaxation of a square distribution to a discrete Maxwellian. The energy drop is at machine precision. The right plot shows the time-history of relative change in entropy. The entropy rapidly increases and remains constant once the distribution function becomes a discrete Maxwellian.}
   \label{fig:sq-relax-cons}
\end{figure}

In this test the relaxation of an initial non-Maxwellian distribution function to a discrete Maxwellian, due to collisions, is studied. The initial distribution function is a step-function in velocity space
\begin{align}
f_0(x,v) = \begin{cases}
1/2v_0 &\qquad |v|<v_0 \\
0 &\qquad |v| \geq v_0,
\end{cases}    
\end{align}
where $v_0 = \sqrt{3} v_{th}$. Piecewise linear and quadratic basis sets on 16 and 8 velocity space cells, respectively, were used and the simulation was run to $\nu t = 5$. In each case the relative change in density and energy are at machine precision showing excellent conservation properties of the scheme. See Fig.\thinspace\ref{fig:sq-relax-cons} in which the time-history of error in normalized energy change is plotted. The errors per step are machine precision. The small drop in energy is due to the damping in the SPP-RK3 scheme. Changing resolution or polynomial has little impact on the magnitude of energy errors, and they always remain close to machine precision. The figure also shows that as the distribution function relaxes, the entropy rapidly increases and then remains constant once the discrete Maxwellian state is obtained. The change in entropy between the $p=1$ and $p=2$ is indicative that different discrete Maxwellians will be obtained depending on grid resolution and polynomial order.  Note that this is not a good test for momentum conservation as the initial momentum is zero.

\begin{figure}
  \incfig{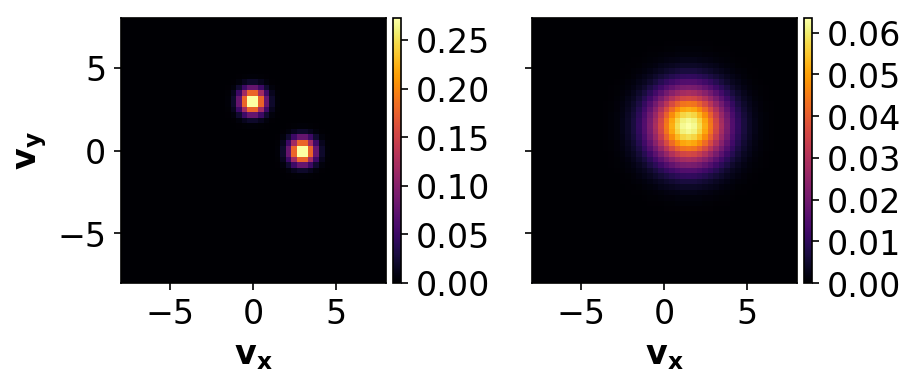}
  \setkeys{Gin}{width=0.45\linewidth,keepaspectratio}
  \incfig{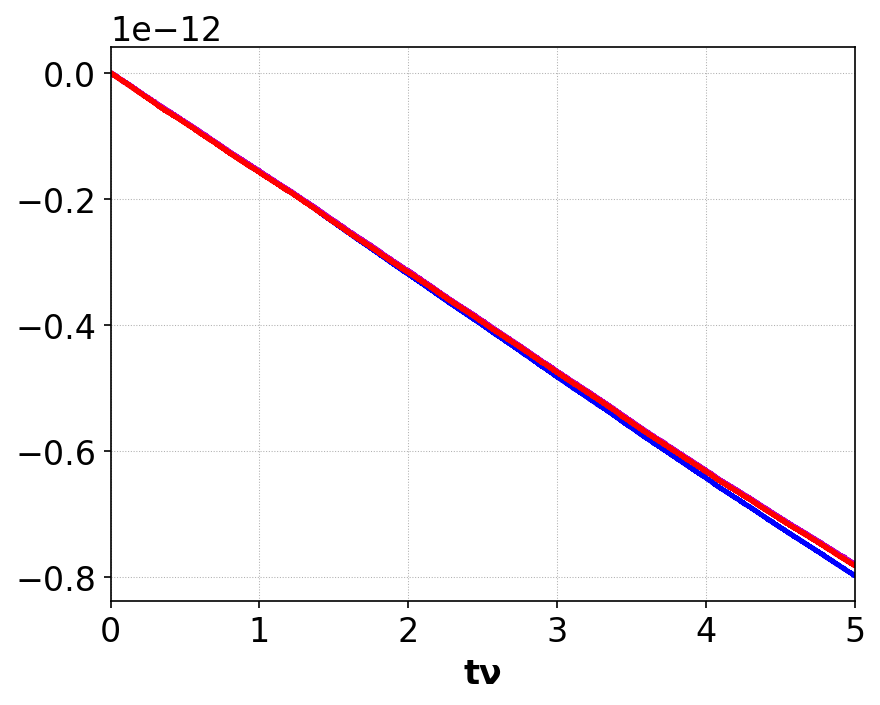}%
  \incfig{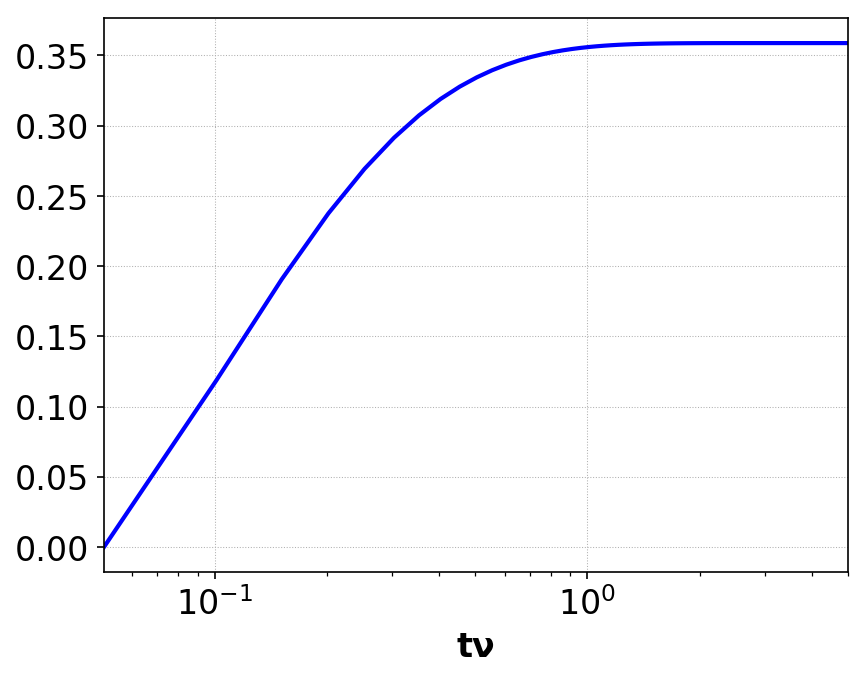}%
  \caption{Initial (top-left), relaxed (top-right) distribution function from non-Maxwellian relaxation test. Conservation  (lower left) of total energy (blue) and momentum (magenta, red) is at machine precision. The entropy (lower-right) increases rapidly and then remains constant once the discrete Maxwellian is obtained.}
   \label{fig:bi-relax-cons}
\end{figure}

Now consider the relaxation in a 1X2V setting. For this the initial condition is selected as a sum of two Maxwellians, the first with drift velocity $\vu = (3,0)$ and the second with drift velocity $\vu = (0,3)$. Both Maxwellians have a thermal spread of $1/2$. A $16\times 16$ grid with $p=2$ Serendipity basis functions is used. As the particles collide the distribution function will relax to a drifting Maxwellian. The simulation is run to $\nu t= 5$. Figure~\ref{fig:bi-relax-cons} shows the initial and final distribution function demonstrating the relaxation to the discrete Maxwellian. The errors in energy and $x$- and $y$-components of momentum are machine precision. Also, the entropy increases monotonically, reaching its steady-state value once the discrete Maxwellian is obtained. These tests demonstrate the high accuracy with which the moments are conserved as well as provide empirical evidence that entropy is a non-decreasing function of time. 

\subsection{Kinetic Sod-Shock problems}

\begin{figure}
  \setkeys{Gin}{width=0.65\linewidth,keepaspectratio}
  \incfig{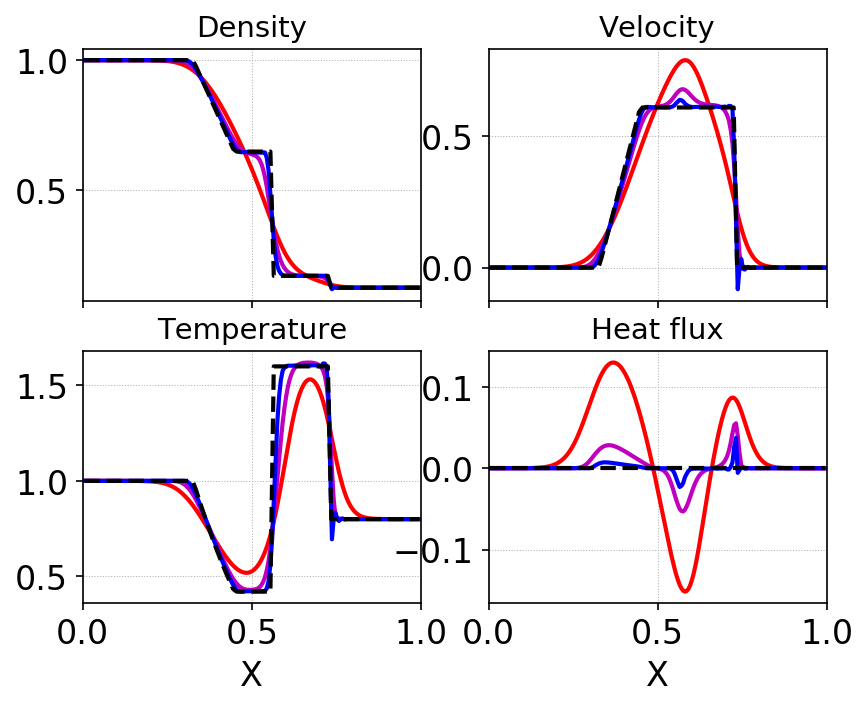}%
   \caption{Density (top-left), velocity (top-right), temperature (bottom-left) and heat-flux (bottom-right) from a Sod-Shock problem. Plotted are results with Knudsen numbers of $1/10$ (red), $1/100$ (magenta) and $1/500$ (blue) with the inviscid Euler results (black dashed) shown for comparison. As the gas becomes more collisional (decreasing Knudsen number) the solutions tend to the Euler result. Note that there is no heat-flux in the inviscid limit.}
   \label{fig:sod-shock-moments}
\end{figure}

Consider a neutral gas. The physics is described by the Boltzmann equation, in which the particles stream freely between collisions. Physically accurate solutions require modelling the collisions as ``hard sphere'' or large-angle collisions. However, the Fokker-Planck operator, modeling small angle collisions, can still be used to test mathematically the convergence of the solution to Euler equations as the mean-free-path is reduced. (In the collisional limit the transport coefficients computed from the full Boltzmann collision operator will differ somewhat from those computed from the FPO. However, in this test we are not concerned with physically accurate transport coefficients but simply a mathematically interesting test problem). In this benchmark, we use the algorithms developed above to study the shock structure in the kinetic regime. For this, we select the classic Sod-shock initial conditions
\begin{align}
  \left[
    \begin{matrix}
      \rho_l \\
      u_l \\
      p_l
    \end{matrix}
  \right]
  = 
  \left[
    \begin{matrix}
      1 \\
      0.0 \\
      1.0
    \end{matrix}
  \right],
  \qquad
  \left[
    \begin{matrix}
      \rho_r \\
      u_r \\
      p_r
    \end{matrix}
  \right]
  = 
  \left[
    \begin{matrix}
      0.125 \\
      0.0 \\
      0.1
    \end{matrix}
  \right]
\end{align}
on a domain $[0,L]$ with the initial discontinuity located at $x=L/2$. Note that for this 1X1V system, the gas adiabatic constant is $\gamma = 3$. This is because the internal energy is $p/(\gamma-1) = \rho v_{th}^2/2$, which means $\gamma=3$. The simulations were run on a $64\times 16$ grid, with piecewise polynomial order 2 elements, $L=1$, and $t_{end}=0.1$. The Knudsen number ($\mathrm{Kn} = \lambda_\textrm{mfp}/L$ and $\nu = v_{th}/\lambda_\textrm{mfp}$) is varied between $1/10$, $1/100$ and $1/500$. In the first case, the gas is collisionless on the time-scale of the simulation, and in the last case, the gas is highly collisional. Hence, in the last case the solution should match, approximately, the solution from Euler equations. 

Figure~\ref{fig:sod-shock-moments} shows the density, velocity, temperature and heat-flux ($q \equiv \langle(v-u)^3 \rangle$ in the notation of \eqr{\ref{eq:mom-def}}) obtained from kinetic simulations. For comparison, the exact solution to the corresponding Euler Riemann problem is also shown. It is observed, as expected, that as the gas becomes more collisional the moments tend to the Euler solution. An interesting aspect of the kinetic results, though, are the viscosity, heat-conductivity and other transport effects. The impact of these are seen in the smoothing out of the shock structures that are sharp in the Euler solution. In particular, the lower-right plot shows the heat-flux, completely absent in the inviscid equations. There is significant heat-flux in the low collisionality case, but this reduces as the collisionality increase.

\begin{figure}
  \incfig{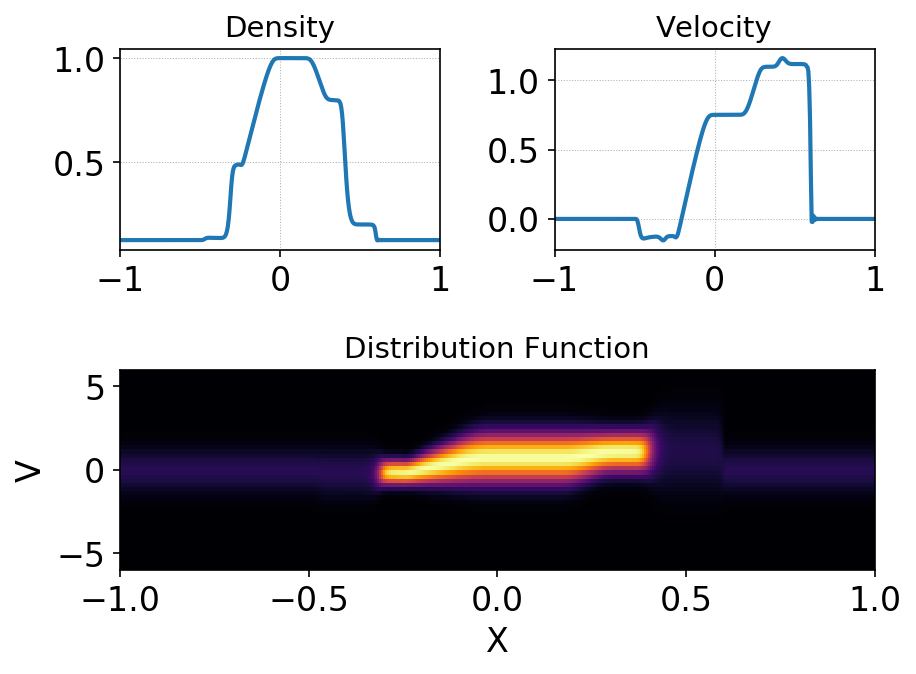}%
   \caption{Density (top-left), velocity (top-right) and distribution function (bottom) for Sod-shock problem with sonic point in rarefaction. Complicated shock structures are formed and are visible both in the moments as well as the distribution function.}
   \label{fig:sod-distf}
\end{figure}

\begin{figure}
  \setkeys{Gin}{width=0.45\linewidth,keepaspectratio}
  \incfig{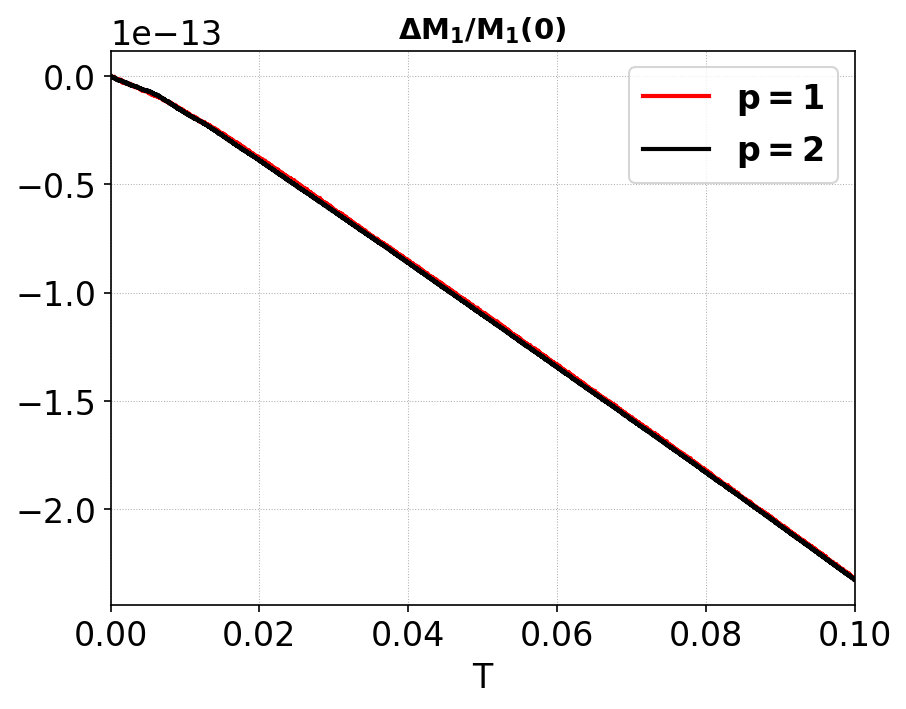}%
   \incfig{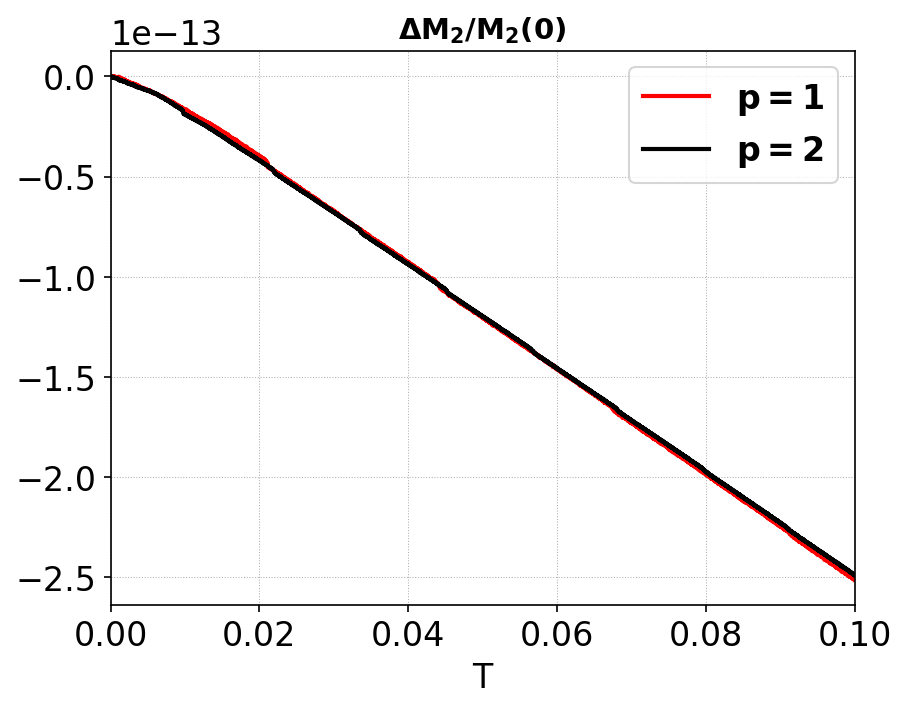}%
   \caption{Relative change in momentum (left) and energy (right) for $p=1$  (red) and $p=2$ (black) cases for Sod-shock problem with sonic point in rarefaction. Momentum and energy drops are at machine precision. The error is nearly independent of polynomial order and only depends on the number of time-steps taken in each simulation.}
   \label{fig:sod-cons}
\end{figure}

We next consider a Sod-shock with a sonic point in the rarefaction wave. The initial conditions are selected as
\begin{align}
  \left[
    \begin{matrix}
      \rho_l \\
      u_l \\
      p_l
    \end{matrix}
  \right]
  = 
  \left[
    \begin{matrix}
      1 \\
      0.75 \\
      1.0
    \end{matrix}
  \right],
  \qquad
  \left[
    \begin{matrix}
      \rho_r \\
      u_r \\
      p_r
    \end{matrix}
  \right]
  = 
  \left[
    \begin{matrix}
      0.125 \\
      0.0 \\
      0.1
    \end{matrix}
  \right].    
\end{align}
In contrast to the standard Sod-shock, this problem is run on a periodic domain $[-1,1]$ with the ``left'' state applied for $|x|<0.3$. The Knudsen number is $1/200$ and the simulation is run to $t=0.1$. As the domain is periodic the total momentum and energy should remain constant, allowing testing conservation properties in more complex setting. Note that the net momentum is not zero in this problem and hence allows testing momentum conservation in a more complex setting than the relaxation test. 

Figure~\ref{fig:sod-distf} shows the density, velocity and distribution function at $t=0.1$. Complex shock structures are visible both in the moments and the distribution function. Figure~\ref{fig:sod-cons} shows the errors in momentum and energy as a function of time for $p=1$ and $p=2$ cases. In each the errors are close to machine precision. Interestingly, the errors seem independent of polynomial order and only depend on the number of time-steps taken in the simulations. These tests demonstrate that even with the streaming terms the conservation properties of the scheme are excellent. Further, as the collisionality is increased we correctly obtain the invscid Euler solutions as expected. 

\subsection{Collisional Landau damping}

\begin{figure}
  \setkeys{Gin}{width=0.4\linewidth,keepaspectratio}
  \incfig{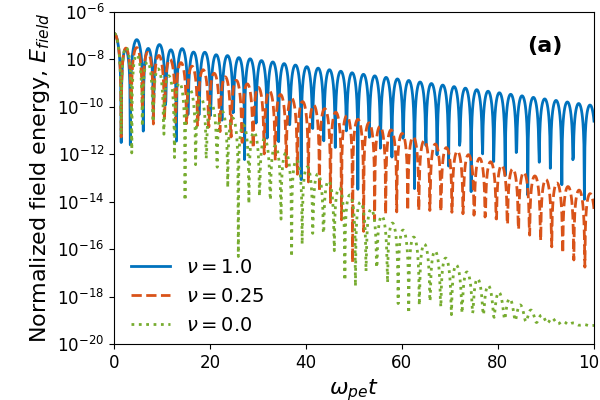}
  \setkeys{Gin}{width=0.525\linewidth,keepaspectratio}
  \incfig{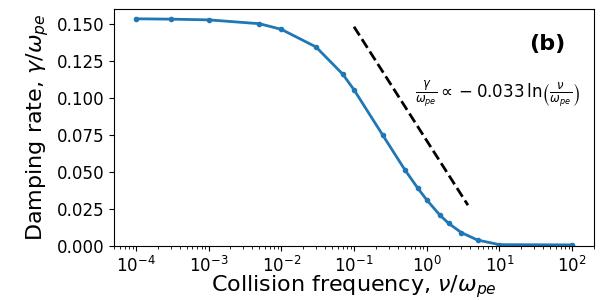}
  \caption{Field energy (left) as a function of time for linear collisional Landau damping problem with varying collisionality. The damping rates (right) are computed as the slope of the decaying electric field energy. The black dashed lined in the right plot shows an analytical estimate of the damping rate computed from expressions found in~\cite{Anderson2007} and agree well with the results computed here.}
  \label{fig:langmuirDamp} 
\end{figure}

Landau damping is the process of damping of electrostatic plasma waves in a collisionless  plasma. It is usually studied in the setting of the collisionless Vlasov-Poisson equations. However, collisions can significantly change the damping rate and in the limit of high collisionality the damping can be ``shut off''. This happens when the mean free path becomes shorter than the wavelength, preventing the particles from resonating with the wave and gaining energy before being scattered via collisions.

In this test the ability of the algorithm to capture the phenomena of collisional Landau damping is shown. The Vlasov-Fokker-Planck equation is coupled to Maxwell's equations to advance the electric field that in turn is used in the Lorentz force to update the distribution function. The electrons are initialized with a perturbed Maxwellian given by
\begin{align}
    f(x,v,0) = \frac{1}{\sqrt{2\pi v_{th}^2}} \exp(-v^2/2v_{th}^2)
    (1+\alpha\cos(kx))
\end{align}
where $k$ is the wave-number,  with $k \lambda_{De} = k v_{te} / \omega_{pe} = 0.5$, and $\alpha$ is the magnitude of the perturbation. Periodic boundary conditions are imposed in the spatial direction. The ion density is set to $n_i(x) = 1$ and is held fixed. To compute the damping rate the decay of the electric field energy is tracked. Figure~\ref{fig:langmuirDamp} shows the electric field energy as a function of time for $\nu = 0, 0.25$ and $1.0$. The collisionless damping rates compare well with exact analytical theory. As the collision frequency increases the damping rate decreases rapidly in the moderate collision frequency limit, as seen in the right panel. Although collisionless Landau damping is very well studied, analytical results in the collisional case are harder to come by. Part of the issue is that in this case the linearized Vlasov-Poisson-Fokker-Planck equation becomes a differential equation in velocity space, needing careful treatment. In~\cite{Anderson2007} a collision operator similar to the one presented here is used to derive analytical estimates of the damping rates as a function of collision frequency, though they consider the 1X3V case and here we are doing the 1X1V case so the results are not expected to exactly match.  Nevertheless, a fit to the slope in the intermediate collisionality transition regime from the theory in~\cite{Anderson2007} (shown in the black dashed line in Fig.\thinspace\ref{fig:langmuirDamp}) shows reasonable agreement with the numerical results here.

\subsection{Heating via magnetic pumping}

\begin{figure}
  \centering
  \begin{subfigure}[b]{0.52\textwidth}
  \includegraphics[width=\textwidth]{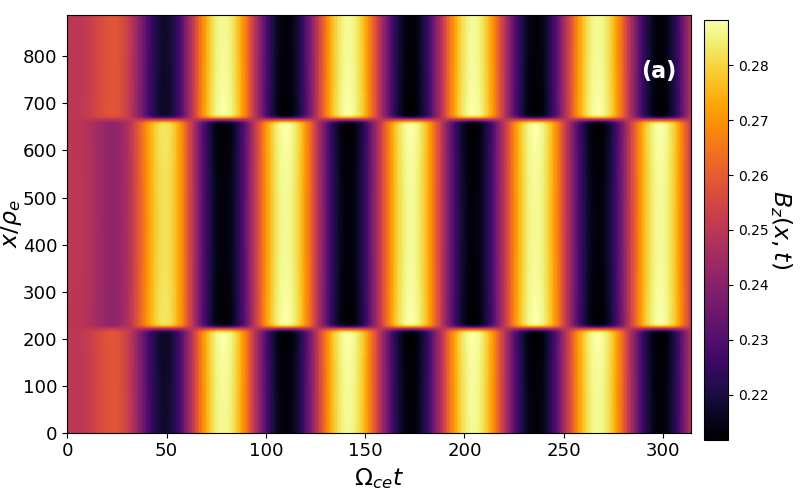}
  \end{subfigure}
  \begin{subfigure}[b]{0.47\textwidth}
  \includegraphics[width=\textwidth]{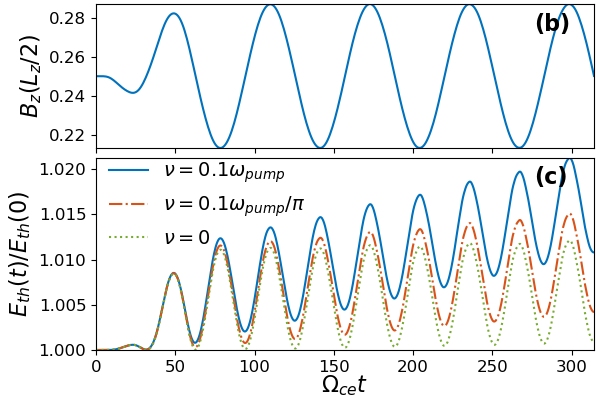}
  \end{subfigure}
  \caption{Time evolution of the magnetic field (left) from the magnetic pumping problem. As the antenna currents ramp up, an oscillating field is created that then transfers energy, via pitch-angle scattering, to the plasma. The top right plots shows the time-history of the field in the middle of the domain and the bottom right figure shows the integrated thermal energy history with various collisionalities. As the collision frequency increases, magnetic pumping is more effective in heating the plasma.}
  \label{fig:magPump} 
\end{figure}

In the final test we combine the FPO with the collisionless Vlasov equation and Maxwell's equations to solve the full Vlasov-Maxwell-Fokker-Planck system of equations to study a specific heating mechanism called magnetic pumping, and an additional viscous mechanism in this parameter regime. In magnetic pumping, oscillations of the magnetic field are converted to particle energy. Magnetic pumping relies on the approximate conservation of magnetic moment, $\mu = m v_\perp^2/2B$ in a strong magnetic field. As the magnetic field increases, to maintain magnetic moment conservation, $v_\perp^2$ should also increase. 
In a collisionless system, if the magnetic field is oscillating slowly compared to the gyro period, then $v_\perp^2$ oscillates up and down in a reversible way and there is no net heating of the plasma.  
However, collisions can provide a route to ``pitch angle scatter'' the energy into the parallel direction, leading to an overall irreversible heating of the plasma. This mechanism was originally proposed by Spitzer in the early days of fusion research, and investigated extensively ~\cite{berger:1958,Laroussi:1989hl}. Recently, this same mechanism was investigated as a potential source of particle heating in the solar wind~\cite{Lichko2017}. We use a similar setup as~\cite{Lichko2017}, with some parameters chosen differently. The study in~\cite{Lichko2017} used a full Coulomb operator evolved with a Monte-Carlo method implemented in a particle-in-cell (PIC) code. Here, we used the simplified form of the Dougherty FPO. Even though this is simpler than the Coulomb operator it contains enough physics to study the mechanism of magnetic pumping, and demonstrate that our algorithm works robustly in a complex plasma problem.

The domain is 4D with one spatial dimension but three velocity dimensions (1X3V) and has extents $[0,200\pi\rho_e] \\ \times[-8v_{th,s}, 8v_{th,s}]^3$ on a $256\times24^3$ grid. Here, $\rho_s = v_{th,s}/\Omega_{cs}$ is the gyroradius of species $s$. A perturbation is driven on a  background magnetic field $\v{B}=\uv{z}\thinspace B_0$ using an antenna that drives currents given by
\begin{align} 
\v{J} = 
\uv{y}J_0
\sin^2\big(0.5\pi\thinspace\min(1, \omega_{\text{ramp}}t)\big) 
\sin(\wpump t)
\left[
\exp\left(-\frac{(x-x_1)^2}{2\sigma_J^2}\right)-
\exp\left(-\frac{(x-x_2)^2}{2\sigma_J^2}\right)
\right]. 
\label{eq:jPump}
\end{align}
(One can think of this antenna as physically realizable as a highly transparent mesh of current carrying wires.)
The current is turned on slowly over one pumping period using $\omega_{\text{ramp}}=\wpump$. This ramping phase ensures that the antenna is ``turned on'' slowly and hence does not excite unwanted waves in the plasma. Further, we need to ensure that the plasma density is low enough that the electromagnetic waves are not ``trapped'' in the density holes that are created around the antenna.

The tests shown here use $\wpump=0.1\omegace$, $x_1=50\pi\rho_e$, $x_2=150\pi\rho_e$, $\sigma_J=200\pi\rho_e/256$ and $\omegace=2.5\omega_{pe}$. We employ a hydrogen mass ratio $m_i/m_e=1836$ and initialize electron and ion species as Maxwellians with zero mean flow, number density $n\thinspace \rho_e^3=2.99\times10^{5}$ and thermal speed $\vte^2/c^2=\beta \omegace^2/[2\omega_{pe}^2(1+\tau)]$. The temperature ratio is $\tau=T_i/T_e=1$ and the ratio between plasma and magnetic pressures is $\beta=2\times10^{-4}$. With these quantities, the normalized background magnetic field amplitude is $\epsilon_0\omega_{pe}B_0/(en)=\omegace/\omega_{pe}$, and we use the normalized driving current density amplitude $J_0/(enc)=\omegace/(2\omega_{pe})$. 

\begin{figure}
 \setkeys{Gin}{width=0.8\linewidth,keepaspectratio}
  \incfig{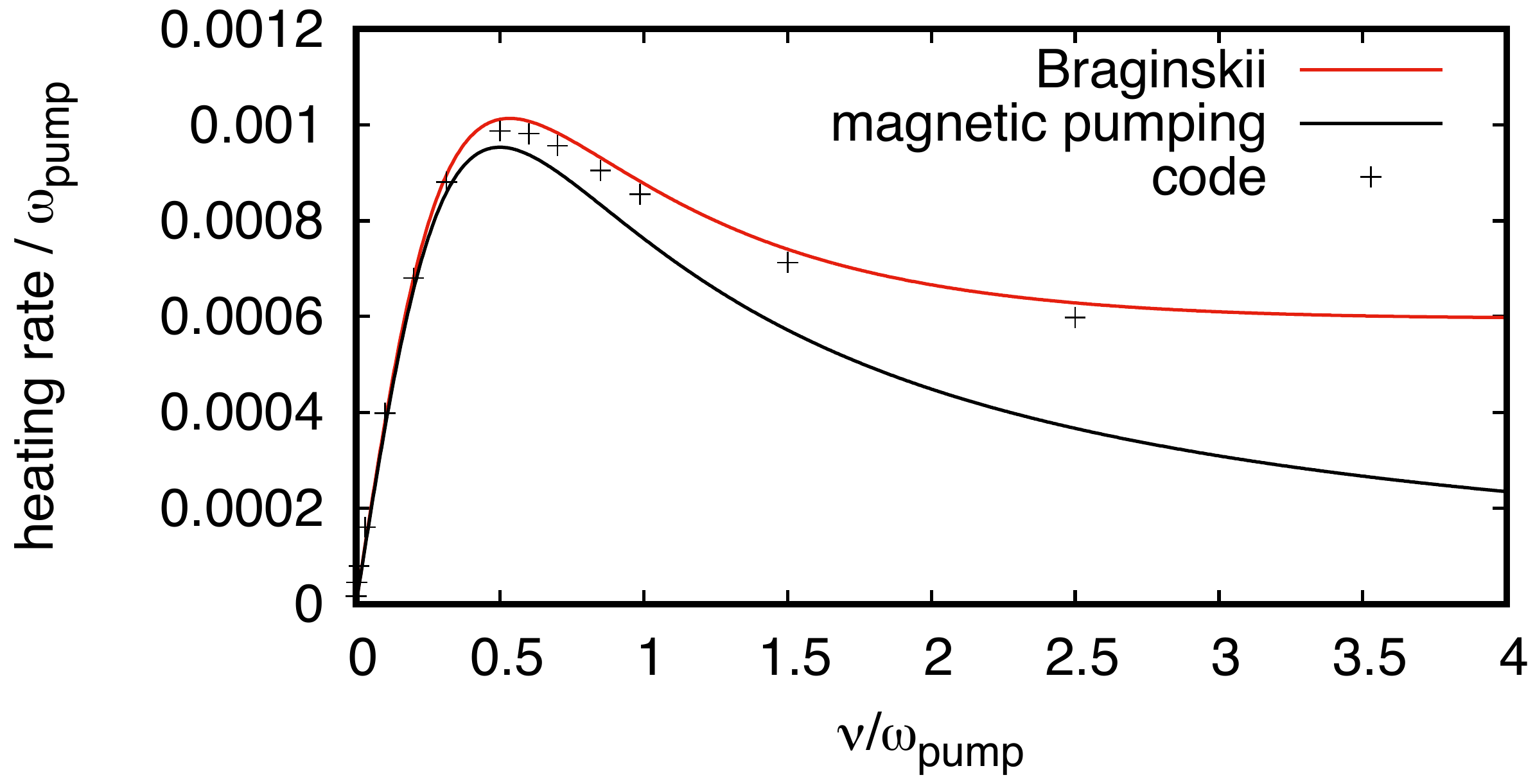}
  \caption{Heating rate via magnetic pumping, and an additional viscous heating mechanism, as a function of normalized collision frequency. The code argees well with magnetic pumping at lower collision frequency, but shows an additional heating mechanism at higher collisionalty due to the viscous damping of out-of-plane flows, which are included in the Braginskii-based theory.}
  \label{fig:heatRates} 
\end{figure}

Figure\thinspace\ref{fig:magPump} shows the evolution of the magnetic field and thermal energy. As the antenna current ramps up an oscillating magnetic field structure is created. The amplitude of oscillations are about $15\%$ of the background. This oscillating energy is then transferred to parallel heating, via pitch angle scattering. This heating is shown in the bottom right panel of the figure which shows that as the collision frequency increases the plasma heating-rate also increases.

Figure\thinspace\ref{fig:heatRates} shows the heating rate as a function of normalized collision frequency. 
The heating rate from standard magnetic pumping theory using the Dougherty FPO is $\gamma_{\rm mp} = (dW/dt )/ W = (2/9) (n_1/n_0)^2 \omega_{pump}^2 \nu / (\omega_{\rm pump}^2 + 4 \nu^2)$, where the central density $n(t) = n_0 + n_1 \sin(\omega_{\rm pump} t)$ (after the initial transients).  
(This heating rate is usually expressed in terms of the relative magnetic field oscillation amplitude, $B_1/B_0$, but that is equal to $n_1/n_0$ if the frozen-in-condition is perfect.  We discuss this further below.)

Interestingly, the code agrees with this magnetic pumping theory at small $\nu/\omega_{\rm pump} \lessapprox 1$, but indicates an additional heating mechanism for larger collisionality.
This trend was also observed, but in a different parameter regime and using Coulomb collisions, in~\cite{Lichko2017}.
There are several differences between this test case and the classic magnetic pumping test problem.  One is that the pump frequency is large compared to the ion gyrofrequency, so the ions are close to stationary as the electrons undergo periodic compressions by the magnetic field.  This means that a large electric field is generated in the $x$ direction, which in turn drives large $y$-directed $E \times B$ flows in the electrons.
We won't go through all of the details of the derivation here, but for $\nu \gg \omega_{\rm pump}$, one can use Braginskii's stress tensor to calculate the viscous heating of the electrons from these flows, finding that Braginskii's heating rate is
$$
\gamma_{B} = \frac{2}{3 n T}\left[ \left( \frac{\eta_0}{3} + \eta_1 \right)\overline{\left( \frac{\partial u_x}{\partial x} \right)^2} + \eta_1 
\overline{\left( \frac{\partial u_y}{\partial x} \right)^2}
\right] 
$$
where $\eta_0 = 0.96 n T \tau_i$ and $\eta_1 = 0.3 n T / (\tau  \Omega_c^2)$ are two of Braginskii's viscosity coefficients.  (These expressions are for $\omega_{\rm pump} \ll \nu \ll \Omega_c$, but are generalized for arbitrary $\nu/\Omega_c$ in Braginskii's article\cite{Braginskii65}.
The $\eta_0$ term is well known to be equivalent to magnetic pumping in the collisional limit\cite{Kulsrud2005,Schekochihin2005}, and asymptotic matching can be done to extend the definition of $\eta_0$ into the low collisionality regime. 
This relates Braginskii's collision time $\tau$ to the Dougherty collision rate by $\tau = 0.52 / \nu$.  The $\eta_1$ term represents additional viscous heating due to classical cross-field momentum transport.

One can calculate the time-averaged squared shearing rate $\overline{\left(\partial u_x / \partial x \right)^2} = (1/2) \omega_{\rm pump}^2 (n_1/n_0)^2$, and $\overline{\left(\partial u_y / \partial x \right)^2} = (1/2)$ $(\omega_{\rm pe}^4 / \Omega_c^2) \ (n_1/n_0)^2$, to find that the out-of-plane flows are actually larger, with $\overline{u_y^2} \approx 2.56 \, \overline{u_x^2}$ for our parameters.  Viscous heating from damping these flows dominates at high collisionality for these parameters.

We have not tried to do a more detailed convergence study for this problem for a few reasons.  One is that Braginskii's transport calculations were based on the full Landau/Rosenbluth FPO, so the ratio of $\eta_1$ to $\eta_0$ will differ some for a transport theory based on the Dougherty operator.  Another is that the some of the above steps assumed that flows $u_x(x,t)$ and $u_y(x,t)$ had a simple triangle wave shape in $x$ so that the plasma was undergoing uniform compression (or expansion) between the antennas, but in reality the compression is not exactly uniform.
The heating rate shown in the figure is measured by computing the slope of the time-averaged temperature near the middle of the simulation domain, $x = 100  \pi \rho_e$, where the amplitude of the density oscillations used in the formulas are also measured.
%
%
%
The finite value of $\omega_{\rm pump} / \Omega_c = 0.1$ should also be accounted for in a more detailed theory.
While $n_1/n_0 = B_1/B_0$ in the ideal frozen-in limit, we find $n_1/n_0 = 0.131$ and $B_1/B_0 = 0.148$, so using $n_1/n_0$ in the formulas gives a slightly smaller heating rate.

Although more could be done in studying this system, this magnetic pumping problem provides an integrated test of the interactions of the various pieces of the code for the Vlasov-Maxwell-Fokker-Planck system.  These types of tests could be repeated in the future when the collision operator is extended to a more complete Rosenbluth form including the velocity-dependence of the collision frequency.

\section{Conclusion}

In this paper we have presented a novel discontinuous Galerkin scheme to solve a class of nonlinear Fokker-Planck equations. Often, this operator is known as the ``Dougherty'' operator or the ``Lenard-Bernstein'' operator.  These are simplified forms of the full Rosenbluth Fokker-Planck collision operator with specific choices of the mean momentum and energy tensor coefficients.  Generalization to the full Rosenbluth collision operator could be considered in future work. Several novel algorithmic innovations are reported here. In particular:
\begin{itemize}\cramplist
    \item The definition of weak-equality and its application to weak-division and construction of the discrete diffusion term is crucial. See Section~\thinspace\ref{sec:weakeq}. This allows an alias-free and consistent approach to computing the primitive moments needed in the Fokker-Planck operator. The definition of weak-equality is also used to describe the reconstruction process needed in the diffusion term.
    \item The use of two integration by parts is required to ensure that momentum and energy, in addition to density, are conserved by the discrete scheme. Without the second integration by parts jumps in the distribution function would be picked up and make the scheme non-conservative. Further, the diffusion term in the FPO requires accounting for finite velocity space extents. See Scheme C, \eqr{\ref{eq:schemec}}, and the conservation proofs in Section~\ref{sec:scheme-theory}. Interestingly, the requirements on momentum and energy conservation, combined with finite-velocity extents, also lead to a unique and consistent method to determine the drift-velocity and thermal speeds. 
    \item For piecewise linear basis sets energy conservation can be obtained by carefully defining the ``star'' moments; see Proposition~\ref{prop:sc_erCons_p1}. Naively, one may have assumed that $\vv^2$ must be included in the basis set. However, with the definition of the star-moments we can continue to conserve energy exactly even when using $p=1$ basis.
\end{itemize}

In addition, we have carefully described the conservation properties of the continuous Fokker-Planck operator. Some of this information is found elsewhere, however, it is useful to include it here, both for completeness and to show the analogous steps required to ensure discrete conservation. In particular, the proof of self-adjointness and the H-theorem shows that the inverse of the Maxwellian is the natural weight to define an inner-product for the FPO. In the numerical scheme we show empirical evidence that the discrete entropy computed from the scheme is a non-decreasing function.

We have tested the scheme with a series of problems of increasing complexity. Simple relaxation tests with just the collision operator show that a square or bi-Maxwellian distribution quickly relaxes to a discrete Maxwellian. In fact, the steady-state solution obtained from evolving just the collision operator \textit{is} the discrete Maxwellian, rather than the projection of the continuous Maxwellian on basis functions. Including the streaming term allows one to study neutral gas dynamics in the long-mean-free path limit. A series of neutral shock problems and the relaxation tests show that momentum and energy are conserved to machine precision. Empirical evidence shows that discrete entropy is non-decreasing. Adding electrostatic terms to the Vlasov equations allowed us to study impact of collisions on Landau damping of Langmuir waves. Finally, the magnetic pumping problem tested the complete Vlasov-Maxwell-Fokker-Planck system, showing that complex collisional kinetic plasma problems can be studied.

The methods presented here are general, and can be extended in a number of ways. First, inter-species collisions are relatively easy to add, although the weak-division calculation of the appropriate intermediate drift velocity and diffusion rates has to be generalized for multi-species energy and momentum conservation. A general formulation is being implemented in \gke\ and will be used to study both gyrokinetic and full-kinetic problems. The proofs of conservation extend to this general case in a straight-forward way. The extension of the scheme to gyrokinetic equations in non-trivial and will be presented in a companion paper. A more challenging task is to extend the scheme to use the Rosenbluth potentials to compute the drag and diffusion coefficients. Rosenbluth potentials are determined by inverting elliptic equations in velocity space. The requirement of momentum and energy conservation will impose constrains on the elliptic solves and their boundary conditions, but these remain to be worked out. Further, to ensure an efficient scheme for the full system, the elliptic solves themselves will require fast methods. This extension forms part of our current algorithm development projects and will be reported on in the future.

\section*{Acknowledgements}

We are grateful for insights from conversations with Petr Cagas, Tess Bernard, Noah Mandell and other members of the \gke\ team. This work used the Extreme Science and Engineering Discovery Environment (XSEDE), which is supported by National Science Foundation grant number ACI-1548562, resources of the Argonne Leadership Computing Facility, which is a DOE Office of Science User Facility supported under Contract DE-AC02-06CH11357, as well as the Discovery cluster at Dartmouth College, the Eddy cluster at the Princeton Plasma Physics Laboratory, and the MIT-PSFC partition of the Engaging cluster at the MGHPCC facility (funded by DoE grant number DE-FG02-91-ER54109). A. Hakim and G. Hammett are supported by the High-Fidelity Boundary Plasma Simulation SciDAC Project, part of the DOE Scientific Discovery Through Advanced Computing (SciDAC) program, through the U.S. Department of Energy contract No. DE-AC02-09CH11466 for the Princeton Plasma Physics Laboratory. A. Hakim is also supported by Air Force Office of Scientific Research under Grant No. FA9550-15-1-0193; M. Francisquez is supported by U.S. Department of Energy grants DOE-SC-0010508 and DE-FC02-08ER54966. J. Juno was supported by a NASA Earth and Space Science Fellowship (Grant No. 80NSSC17K0428).

.

\bibliographystyle{elsarticle-num} 
\bibliography{references}

\appendix
\section*{Getting \gke\ and reproducing the results}

To allow interested readers to reproduce our results and also use \gke\ for their applications, in this Appendix we provide instructions to get the code (in both binary and source format) as well as input files used here. Full installation instructions for \gke\ are provided on the \gke\ website (\url{http://gkeyll.readthedocs.io}). The code can be installed on Unix-like operating systems (including Mac OS and Windows using the Windows Subsystem for Linux) either by installing the pre-built binaries using the conda package manager (\url{https://www.anaconda.com}) or building the code via sources. The input files used here are under version control and can be obtained from the repository at \url{http://bitbucket.org/ammarhakim/d-fpo-paper-inp-files}. Except for the magnetic pumping problem, all other tests can be run on a laptop.


\end{document}